\documentclass[9pt]{article}

\usepackage{preamble}
  
\title{\Large \textbf{A visual introduction to information theory}}

\date{} 

\author[1,2]{Henry Pinkard}
\author[1]{Laura Waller}

\affil[1]{Department of Electrical Engineering and Computer Sciences, University of California, Berkeley, CA, USA}
\affil[2]{Computational Biology Graduate Group, University of California, Berkeley}

\affil[*]{\normalfont{\textbf{Correspondence: hbp [at] berkeley.edu}}}

\begin{document}

\maketitle

\renewcommand{\abstractname}{\vspace{-26pt }} 

\begin{abstract} \bf

Information theory, though originally developed for communications engineering, provides mathematical tools with broad applications across science. These tools characterize the fundamental limits of data compression and transmission in the presence of noise. Here, we present a visual, intuition-driven guide to key concepts in information theory. We show how entropy, mutual information, and channel capacity follow from basic probability, and how they determine the shortest possible encoding of a data source and the maximum rate of reliable communication through a noisy channel. Our presentation assumes only a familiarity with basic probability theory.

\end{abstract}

\section{Introduction}

In the 1940s, Claude Shannon \cite{Shannon1948} gave information a precise mathematical definition grounded in probability. Since then, information theory has established the fundamental limits of data compression and reliable transmission across noisy networks, forming a foundation for the digital world. Its mathematical tools have found applications in statistics, machine learning, cryptography, quantum computing, biology, and many other areas.

Here we introduce the foundations of information theory, with an emphasis on intuition. This is a concise introduction to key ideas; the textbooks \cite{MacKay2005, Cover2011} and Shannon's original formulation \cite{Shannon1948} offer a more comprehensive treatment.

\paragraph{Main ideas}

Information theory made a familiar intuition mathematically precise: \textit{information} means \textit{knowledge about something unknown}. Shannon described information as something that ``can be treated very much like a physical quantity such as mass or energy."

Shannon defined information in terms of probability: the probability distribution over potential messages is the \textit{only} thing that matters for transmitting information. The actual content of the messages is irrelevant. The randomness of a sequence of colored marbles is described the same way as the randomness of a string of letters in English text or a sequence of pixels in images from the Hubble telescope.

This paper covers the two key problems in information theory. The first is \textbf{source coding} (data compression): how succinctly can we record a sequence of random events, on average? This subdivides into \textbf{lossless compression}, where the original sequence must be perfectly reconstructed, and \textbf{lossy compression}, where some distortion is tolerated to achieve a smaller encoding.

The second is \textbf{channel coding} (data transmission): given a \textbf{noisy channel} that introduces distortions, how can we encode a sequence so that the original can be recovered? This too subdivides into perfect transmission and transmission with distortions.

\subsection{Notation}

\begin{center}
\fbox{\parbox{1\textwidth}{%
\begin{tabular}{ l l }
 $\xrv$ & a random variable, which will also be called a ``random event"  \\ 
 $x$ & a particular outcome of $\xrv$  \\  
 $\mathcal{X}$ & the outcome space/set of possible outcomes for $\xrv$. $x \in \mathcal{X}$ \\
 $| \mathcal{X} |$ & the number of possible outcomes in the outcome space \\
 $\prob{\xrv}(x)$, $p(\xrv=x)$ or $p(x)$ & the probability that the random event $\xrv$ has outcome $x$ \\
 $H(\xrv)$ & the entropy of $\xrv$ \\
 $\maxentropy(\mathcal{X})$ & the maximum entropy for the state space $\mathcal{X}$ \\
 $W(\xrv)$ & the redundancy of $\xrv$ \\
 $\mathbf{X} = \mathrm{X_1}, \mathrm{X_2},...$ & a stochastic process: an ordered sequence of random variables \\
 $\mathbf{\mathcal{X}^2} = \mathcal{X}\times\mathcal{X}$ & the state space for ($x_1$, $x_2$) tuples where $x_1 \in \mathcal{X}, x_2 \in \mathcal{X}$ \\
 $\mathcal{P}_{\mathcal{X}}$ & The set of probability distributions on the space $\mathcal{X}$ \\
 $\mathbf{p}_{\xrv}$ & A vector representation of the probabilities of the distribution of $\xrv$ \\
  $\mathbf{P}_{\xrv, \yrv}$ & A matrix representation of the joint distribution of $\xrv$ and $\yrv$ \\
  $\mathbf{P}_{\yrv \mid \xrv}$ & A matrix representation of conditional distribution $\yrv$ given $\xrv$ \\
  $C$ & The channel capacity: \\ &the maximum information that can be transmitted by a channel
\end{tabular}
}}
\end{center}

\section{Core quantities: information, entropy, and mutual information}

Information theory rests on a small set of quantities that measure uncertainty and the relationships between random events. Entropy measures how uncertain a single random event is on average, and determines the limits of data compression. Mutual information measures how much observing one random event reduces uncertainty about another, and determines the limits of data transmission. Together with their extensions to stochastic processes, continuous distributions, and lossy compression, these quantities provide the foundation for the rest of the paper.

\subsection{What is information?}
\label{what_is_info_sec}

Gaining information and reducing uncertainty about a random event are the same thing. What will the weather be like in 3 hours? It's uncertain. But from now until then, we can continuously acquire information by observing the current weather, and 3 hours from now our uncertainty will have collapsed to zero. It no longer appears random to us.

\begin{figure*}[htbp]
\centering
\includegraphics[width=\linewidth]{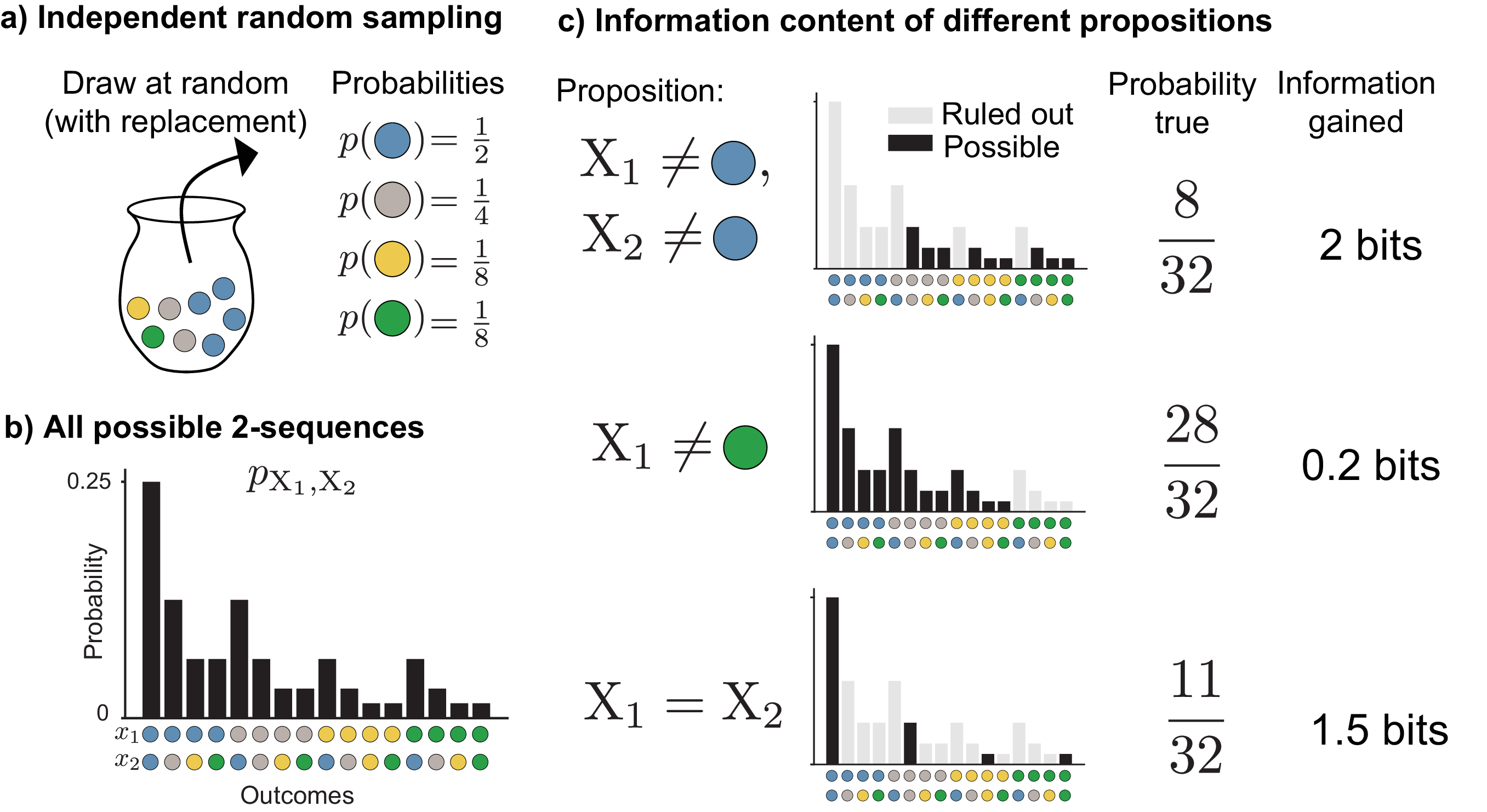}
\caption{\textbf{Equivalence of probability and information}  \textbf{a)} A sequence of two marbles is drawn at random (with replacement) from an urn, giving rise to \textbf{b)} a probability distribution over the 16 possible two-color sequences. \textbf{c)} Learning that a proposition about the two colors drawn is true enables the elimination of certain outcomes. For example, learning neither marble is blue eliminates 7 of the 16 possible outcomes, containing $\frac{3}{4}$ of the probability mass. Eliminating probability mass, reducing uncertainty about the outcome, and gaining information are all mathematically equivalent. Reduction of 50\% of the probability mass corresponds to 1 bit of information
}
\label{what_is_information}
\end{figure*}

The information content of a random event depends only on its probability distribution, not on what the event represents. Because of this, any source of random events can stand in for any other. We use colored marbles as a recurring illustration: drawing a marble at random from an urn containing $\texttt{\color{blue}{blue}}$, $\texttt{\color{OliveGreen}{green}}$, $\texttt{\color{Dandelion}{yellow}}$, and $\texttt{\color{gray}{gray}}$ marbles \figrefsub{what_is_information}{a}.

Formally, we have a finite, discrete set of outcomes, a probability distribution over them, and a sequence of independent and identically distributed (IID) draws. The concepts introduced here generalize to the non-IID case (Sec. \ref{entropy_rate_sec}) and continuous variables and probability density functions (Sec. \ref{differential_entropy_sec}).

More common outcomes provide us with less information, and rarer outcomes provide us with more. To understand why, suppose someone has drawn two colored marbles from an urn, but we don't know what they've drawn. We'll denote these two random events with $\xrv_1$ and $\xrv_2$. Their joint probability distribution $\prob{\xrv_1, \xrv_2}$ tells us the probability of each of the 16 possible two-color sequences \figrefsub{what_is_information}{b}. Suppose we then learn some facts about what was drawn (but not the exact outcome). For example, we might learn that neither marble is $\texttt{\color{blue}{blue}}$, or the first marble isn't $\texttt{\color{OliveGreen}{green}}$, or the two marbles are different colors \figrefsub{what_is_information}{c}.

Learning each fact allows us to rule out possibilities for the outcome of $\xrv_1, \xrv_2$. The less often the fact is true, the more possible outcomes we can rule out. For example, for the distribution shown in \textbf{Figure \ref{what_is_information}}, with probabilities $\frac{1}{2}$, $\frac{1}{8}$, $\frac{1}{8}$, and $\frac{1}{4}$ for $\texttt{\color{blue}{blue}}$, $\texttt{\color{OliveGreen}{green}}$, $\texttt{\color{Dandelion}{yellow}}$, and $\texttt{\color{gray}{gray}}$, knowing the two marbles are the same color rules out $12$ of the $16$ possible outcomes, eliminating $\frac{21}{32}$ of the probability mass. The more possibilities we can rule out, the more information we've gained. Rarer outcomes contain more information.

We can calculate exactly how much information an outcome provides from its probability. By convention, we measure in base-2 logarithms, giving the \textbf{bit} as the unit of information: each time we rule out half of the probability mass, we gain exactly $1$ bit. The information in an outcome with probability $p$ is:

\begin{align*}
\pointinformation    \label{point_information_eqn}
\end{align*}

where the base 2 will be omitted in subsequent sections. In our marble example, if we learn that neither marble is $\texttt{\color{blue}{blue}}$, we're left with only $\frac{1}{4}$ of the probability mass, and we have gained $2$ bits of information, since we have halved the probability mass twice.

The word ``bit" also describes a container that can hold information: a binary digit that is either \texttt{0} or \texttt{1}. A container can hold less than 1 bit of information, in the same way a 1 liter bottle can hold only $\frac{1}{2}$ liter of water. Unless stated otherwise, ``bit" here refers to the unit of information.

The probability-weighted average of the information across all outcomes is the \textbf{entropy}, denoted $H(\xrv)$:

$$H(\xrv) = \sum_{x \in \mathcal{X}} p(x) \log \frac{1}{p(x)} $$

Entropy can be interpreted as how surprising a random event is on average: more random events are more surprising, and observing their outcome yields more information. Conversely, higher entropy means more uncertainty to begin with, so we must acquire more information to be certain about the outcome.

Entropy is maximized when probability is spread as evenly as possible over outcomes, not when there are many extremely rare outcomes. Although rarer outcomes individually contain more information, their decreasing probability outweighs their increasing per-outcome information: the contribution of each outcome to entropy, $p \cdot \log \frac{1}{p}$, goes to zero as $p \rightarrow 0$ because $p$ decreases linearly while $\log \frac{1}{p}$ increases only logarithmically. Section \ref{redundancy_sec} discusses maximum entropy distributions in more detail.

\subsection{Entropy and data compression}

Entropy determines the shortest possible binary encoding of a sequence of random events, a problem known as \textbf{source coding}. To see why, consider drawing a sequence of colored marbles from an urn and recording each outcome using binary \textbf{codewords}, such as \texttt{1}, \texttt{10}, etc. Our goal is to choose an encoding scheme that will (on average) yield the shortest binary description of our sequence without introducing any ambiguity as to what the outcomes were. This is a \textbf{lossless compression} problem.

\begin{figure*}[htbp]
\centering
\includegraphics[width=\linewidth]{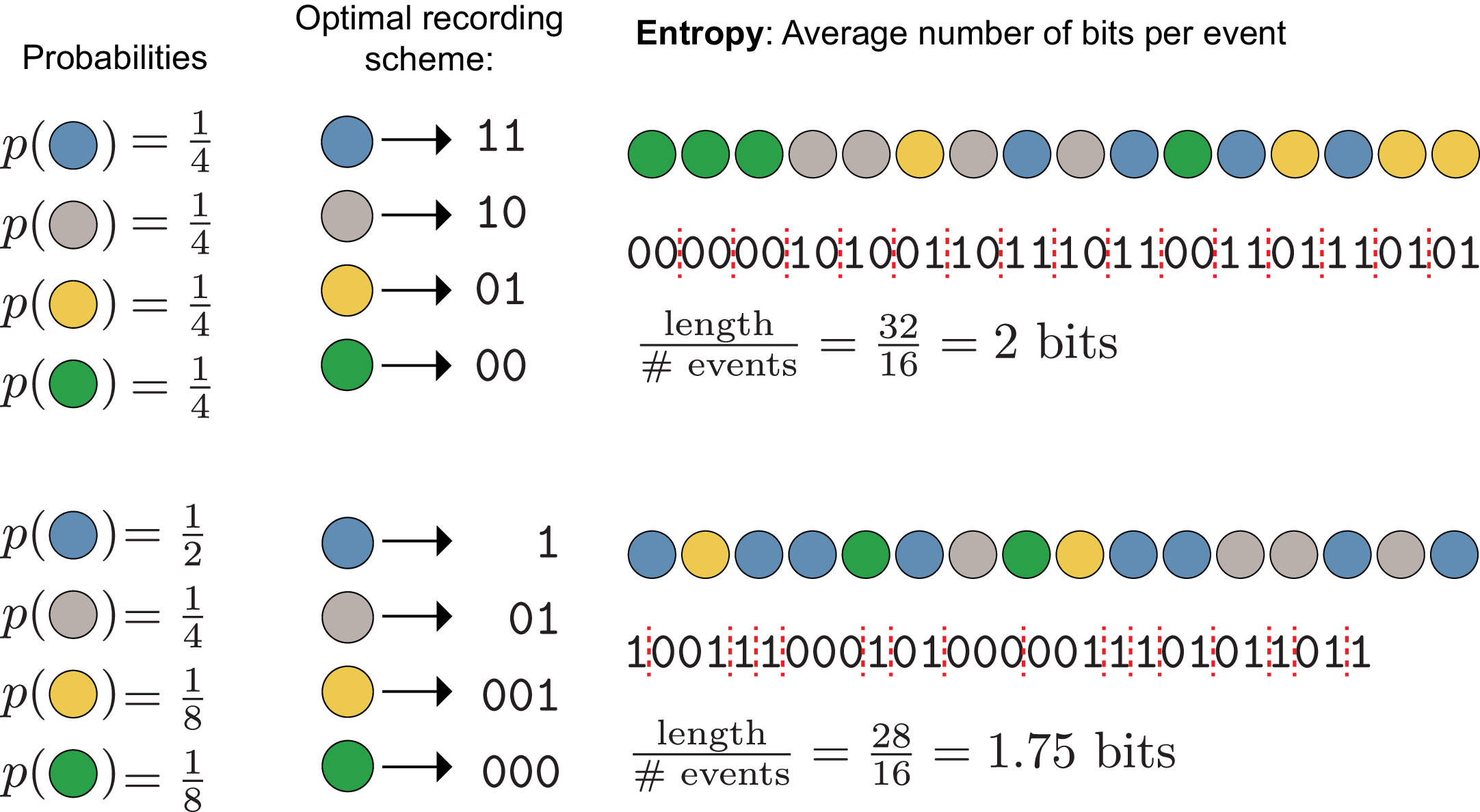}
\caption{\textbf{Entropy} can be interpreted as the average length of the shortest encoding of a sequence of random events, which here are repeated draws (with replacement) of colored marbles from an urn. \textbf{(Top)} With equal probability of each color, the shortest binary recording assigns a two-digit binary string code to each event. The entropy is the average number of bits per event of a typical sequence: 2 bits. \textbf{(Bottom)} When some colors are more likely than others, the more probable ones can be recorded as shorter binary strings to save space. This gives a smaller entropy: 1.75 bits.}
\label{what_is_entropy}
\end{figure*}

Both the information content and the optimal encoding depend on the probabilities of each outcome. For example, consider a random variable $\xrv$ on the \textbf{outcome space} $\mathcal{X}$ with associated probabilities $p_{\xrv}$ \figrefsub{what_is_entropy}{, top}:

\begin{align*}
  \mathcal{X} = \{\tt{\color{blue}{blue}}, \tt{\color{gray}{gray}},
  \tt{\color{Dandelion}{yellow}},
  \tt{\color{OliveGreen}{green}}   \} \\
  p_{\xrv} = \{\tfrac{1}{4}, \tfrac{1}{4}, \tfrac{1}{4}, \tfrac{1}{4} \}
\end{align*}

The shortest lossless encoding scheme given these probabilities is a unique 2 digit binary codeword for each outcome: $\texttt{00}$, $\texttt{01}$, $\texttt{10}$, $\texttt{11}$, because the average information in each random event (its \textbf{entropy}) is 2 bits.

Alternatively, consider the case where the outcomes are not equally probable \figrefsub{what_is_entropy}{, bottom}, but instead:

\begin{align*}
  \mathcal{X} = \{\tt{\color{blue}{blue}}, \tt{\color{gray}{gray}},
  \tt{\color{Dandelion}{yellow}},
  \tt{\color{OliveGreen}{green}}   \} \\
  p_{\xrv} = \{\tfrac{1}{2}, \tfrac{1}{4}, \tfrac{1}{8}, \tfrac{1}{8} \}
\end{align*}

Now the outcome is more predictable: it is more likely to be $\texttt{\color{blue}{blue}}$ and less likely to be $\texttt{\color{Dandelion}{yellow}}$ or $\texttt{\color{OliveGreen}{green}}$. We are more certain, so on average we should need less information to describe each draw.

We can shorten the average encoding by assigning shorter codewords to more probable outcomes and longer ones to less probable outcomes. The shortest lossless encoding scheme uses the codewords $\texttt{1}$, $\texttt{01}$, $\texttt{001}$, $\texttt{000}$ (this encoding is not unique; we could always swap $\texttt{0}$s and $\texttt{1}$s). This is a \textbf{prefix code}: no codeword is a prefix of another, so codewords can be uniquely decoded even when concatenated. Dividing the total encoding length by the number of events, the average information (entropy) per draw is $1.75$ bits.

If we instead used the 2 digit binary code, the average encoding would be longer than necessary, introducing \textbf{redundancy}.

Entropy also describes the uncertainty about the outcome of a random variable. If $p(\tt{\color{blue}{blue}}) = 1$ and all other probabilities equal $0$, the entropy is $0$. In this case there is no uncertainty because we are sure of the outcome even before seeing it.

\subsection{Redundancy}
\label{redundancy_sec}

\textbf{Redundancy} measures the gap between a random event's entropy and the maximum possible entropy on its outcome space. Entropy is maximized when all outcomes are equally probable, the \textbf{maximum entropy} distribution $\maxentropy(\mathcal{X})$ \figrefsub{what_is_entropy}{, top}. Equal probabilities yield the most uncertain outcome and the longest possible optimal binary encoding.

When all outcomes are equally probable, maximum entropy equals the logarithm of the number of possible outcomes:

$$\maxentropy = \log_2{| \mathcal{X} |}$$

where $| \mathcal{X} |$ is the number of elements in $\mathcal{X}$. More possible outcomes means more potential uncertainty.

The \textbf{redundancy} $W$ is the difference between this maximum entropy and the actual entropy of the random variable:

\begin{align*}
    W(\xrv) = \maxentropy(\mathcal{X}) - H(\xrv)
\end{align*}

\subsection{General data compression limits}
\label{source_coding_limit}
The examples in \textbf{Figure \ref{what_is_entropy}} rely on two simplifications that do not hold in general.

First, we've chosen the probabilities to be all of the form $\frac{1}{2^{k}}$ where $k$ is a positive integer. This makes the information in each event equal to an integer number of bits, which means we can fully compress the sequence by encoding each event individually. If we hadn't done this, the information content of a single event might equal a fractional number of bits, and encoding in a one digit binary string would waste a fraction of a bit of space and yield a redundant binary representation. This is why entropy is in general a lower bound on the expected length of the shortest binary encoding, rather than always exactly equal. We can approach this bound more closely by encoding multiple events together into a single binary string, known as \textbf{block encoding}.

Second, the sequences shown contain exactly the expected number of each color. For example, the bottom sequence is half ($\frac{8}{16}$ events) $\tt{\color{blue}{blue}}$, a quarter ($\frac{4}{16}$ events) $\tt{\color{gray}{gray}}$, etc. In reality, different random sequences can have very different encoding lengths. For example, a sequence of all $\tt{\color{blue}{blue}}$ (the most probable sequence) would have a 16-bit encoding, while a sequence of all $\tt{\color{OliveGreen}{green}}$ (one of the least probable) would have a 48-bit encoding. These sequences in \textbf{Figure \ref{what_is_entropy}} are called \textbf{typical sequences}.

\subsubsection{Typical sequences}
\label{typicality_sec}

A \textbf{typical sequence} is an outcome whose information content is close to the entropy of the random event to which it belongs. The \textbf{typical set} is the set of all typical sequences, usually a small subset of all possible sequences. For a sequence of $N$ independent and identically distributed random variables $\xrv_1, \xrv_2,...,\xrv_N$, the average information content is $H(\xrv_1) + H(\xrv_2) +... + H(\xrv_N) = NH(\xrv)$. For a small positive number $\epsilon$, a typical sequence is one that satisfies:

\begin{align*}
    H(\xrv) - \epsilon \leq -\frac{1}{N} \log p(x_1,x_2,...,x_N)\leq H(\xrv) + \epsilon
\end{align*}

\begin{figure*}[htbp]
\centering
\includegraphics[width=\linewidth]{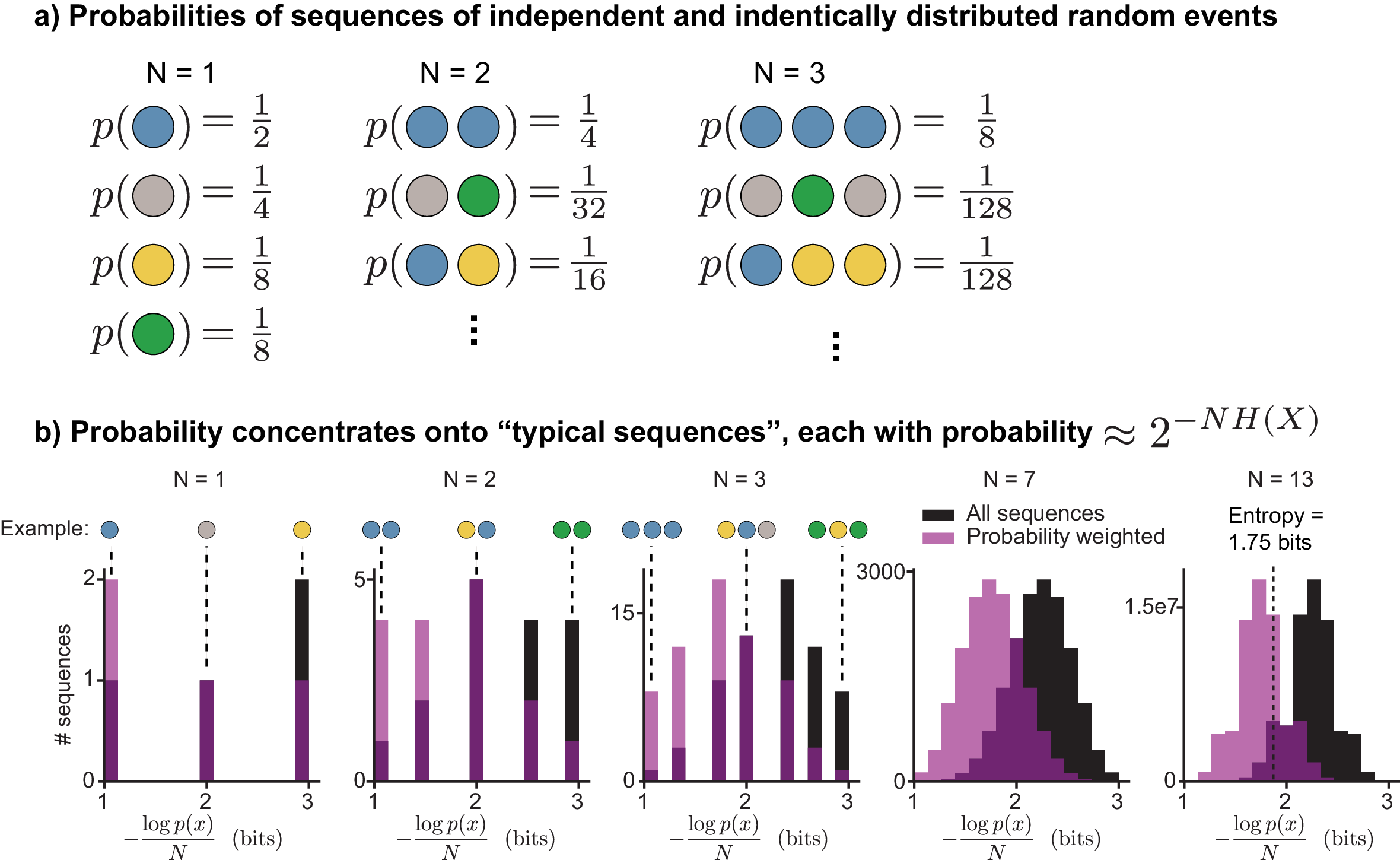}
\caption{\textbf{Typical sequences} \textbf{a)} Example sequences of independent and identically distributed events with increasing length ($N$). \textbf{b)} Histograms of the information (i.e. $-\frac{\log p(x)}{N}$) of each possible sequence with length $N$. Black shows the histogram of every possible sequence. Magenta shows the distribution of probability-weighted sequences (i.e. the expected distribution one would get by taking a random sample). As $N$ increases, nearly all of the probability mass concentrates on a tiny subset of the total number of sequences: typical sequences. There are $\approx 2^{NH(\xrv)}$ typical sequences each with probability $\approx 2^{-NH(\xrv)}$.}
\label{typicality_fig}
\end{figure*}

Does the choice of $\epsilon$ matter? For the sequences in \textbf{Figure \ref{what_is_entropy}}, we've conveniently avoided this question since the sequences shown have an information content exactly equal to $NH(\xrv)$. But as $N \rightarrow \infty$, the choice becomes irrelevant, because all of the probability mass concentrates on the typical set defined with \textit{any positive value of $\epsilon$} \figref{typicality_fig}. All typical sequences occur with approximately the same probability: $\approx 2^{-NH(\xrv)}$.

Since the total probability is $1$ and each typical sequence occurs with probability $\approx 2^{-NH(\xrv)}$, there must be $\approx 2^{NH(\xrv)}$ typical sequences. This result is the \textbf{asymptotic equipartition property (AEP)}, a direct consequence of the Law of Large Numbers. It underlies the proofs of the fundamental limits of both data compression and data transmission.

In short, the AEP says that as sequence length grows, the probability on non-typical sequences vanishes, so we can focus on the typical set of $\approx 2^{NH(\xrv)}$ sequences, each occurring with probability $\approx 2^{-NH(\xrv)}$.

Lossless compression of a length-$N$ sequence requires $NH(\xrv)$ bits. A compression scheme encoding only the typical set would need $\log_2 2^{NH(\xrv)} = NH(\xrv)$ bits for unique binary strings. Since negligible probability mass falls outside the typical set, this scheme is lossless as $N \rightarrow \infty$. Fewer bits would force different typical sequences to map to the same binary string. More bits would waste space on non-typical sequences that occur with nearly zero probability.

The typical set is useful for theoretical work because the number of sequences it contains can be counted, but (for non-uniform distributions) it does \textit{not} include the most probable individual sequences. For example, in \textbf{Figure \ref{typicality_fig}b}, the sequence for each $N$ that consists of $N$ $\tt{\color{blue}{blue}}$s in a row is the most probable sequence, but would be omitted from the typical set for a small value of $\epsilon$. For finite $N$, better compression can be achieved by assigning the shortest codewords to the \textit{most probable} sequences rather than to typical ones. For example, the all-$\tt{\color{blue}{blue}}$ sequence is more probable than any typical sequence, so giving it a shorter codeword would reduce the average encoding length. As $N \rightarrow \infty$, this becomes less and less true, because the most probable sequences contain vanishingly small total probability mass.

\begin{figure*}[htbp]
\centering
\includegraphics[width=\linewidth]{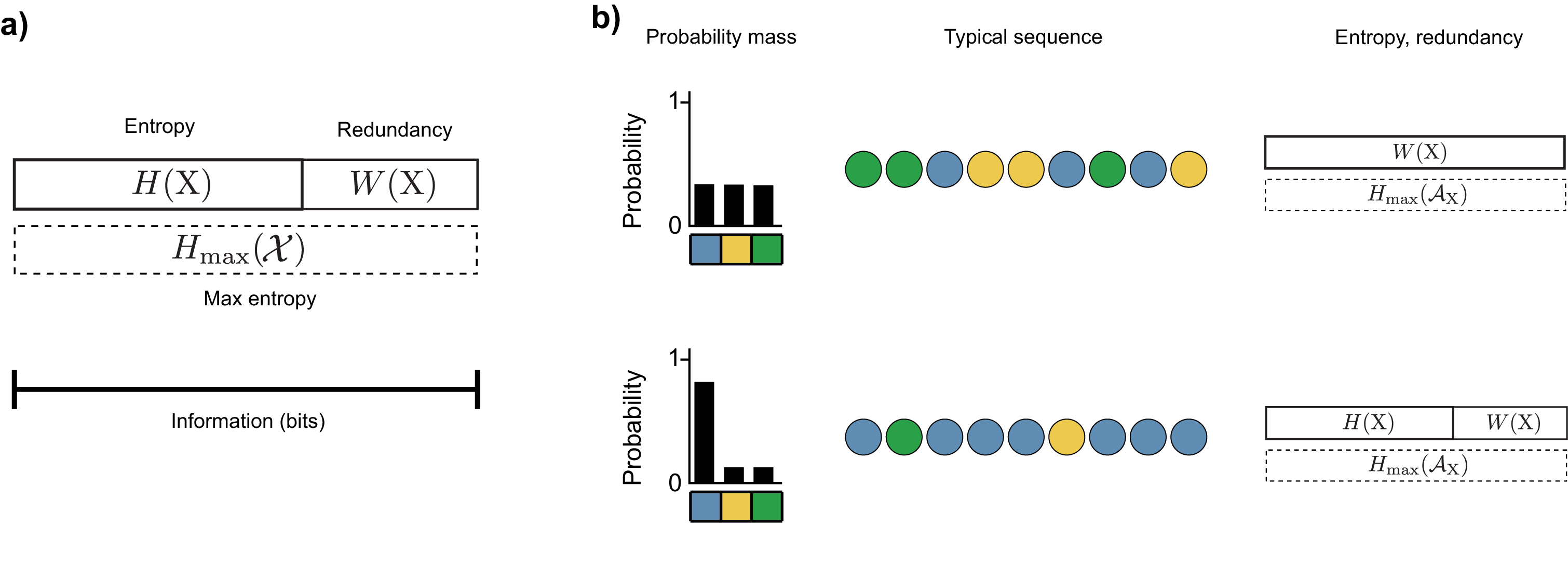}
\caption{\textbf{Probability, redundancy, and typicality}. \textbf{a)} The redundancy of a random variable $\xrv$ is equal to the difference between its entropy $H(\xrv)$ and the maximum possible entropy on its outcome space $\maxentropy(\mathcal{X})$. \textbf{b)} Distributions with more concentrated probability mass have higher redundancy. (Top) The equal probability case, (Bottom) the concentrated probability case. (Left) Probability distribution over a single event of an independent and identically distributed sequence, (Middle) a typical sequence of events from this distribution. (Right) The entropy, redundancy and maximum entropy. }
\label{redundancy_fig}
\end{figure*}

\textbf{Figure \ref{redundancy_fig}} summarizes the relationship between entropy, redundancy, probability distributions, and typical sequences.

\subsection{Mutual information}
\label{mutual_information_intro}

\textbf{Mutual information} quantifies how much knowing the outcome of one random event reduces uncertainty about another. It provides the basis for reasoning about information transmission.

Consider the scenario of drawing not colored marbles, but colored objects from an urn. We have the same four possible colors, $\texttt{\color{blue}{blue}}$, $\texttt{\color{OliveGreen}{green}}$, $\texttt{\color{Dandelion}{yellow}}$, and $\texttt{\color{gray}{gray}}$, and now there are also four possible object shapes: $\blacklozenge$, $\blacktriangle$, $\bigstar$, and $\bigbullet$. Our random draws will now be shape-color combinations like $\color{blue}{\blacktriangle}$, $\color{OliveGreen}{\blacklozenge}$, $\color{Dandelion}{\blacklozenge}$, etc. $\xrv$ will still represent the object's color, and now $\yrv$ will represent its shape. The joint probability distribution over $\mathcal{X} \times \mathcal{Y}$ governs which shape-color combinations occur.

Suppose that we do not observe which object was drawn, but only a black and white photograph in which all colors look the same. What can we say about the object's color ($\xrv$), having only observed its shape $\yrv$? In other words, how much information does shape convey about color? The answer is given by the \textbf{mutual information} between $\xrv$ and $\yrv$.

Mutual information, denoted $I(\xrv; \yrv)$, quantifies how much observing one random event $\yrv$ reduces our uncertainty about another $\xrv$. The minimum possible value is $I(\xrv; \yrv)=0$, which means that $\xrv$ and $\yrv$ are independent: observing shape does not reduce uncertainty about color at all. The maximum possible value is $\min(H(\xrv), H(\yrv))$, because one event cannot convey more information about another than either contains on its own. Mutual information is symmetric: $I(\xrv; \yrv) = I(\yrv; \xrv)$.

Another common way of quantifying dependence between two variables is through correlation. Unlike correlation, which captures only linear relationships, mutual information captures any statistical dependency between two variables.

To compute mutual information, we need the joint probability distribution of $\xrv$ and $\yrv$, i.e., which shapes come in which colors. For a particular draw, say $\color{blue}{\bigstar}$, the information that color and shape convey about one another is the \textbf{point-wise mutual information}:

\begin{align*}
    \log \left( \frac{p(x, y)}{p(x)p(y)} \right)
\end{align*}

Point-wise mutual information is analogous to the information of a single outcome. It can be rewritten in a few different ways, which allow for different interpretations. For example:

\begin{align*}
    \log \left( \frac{p(x, y)}{p(x)p(y)} \right) 
    &= \log \left( \frac{p(x \mid y)}{p(x)} \right)  \\
    &= \log p(x \mid y) - \log p(x) \\
    &=  \log \frac{1}{p(x)} - \log \frac{1}{p(x \mid y)} 
\end{align*}

The final line can be interpreted as the surprise of the outcome $x$ minus the surprise of $x$ when $y$ is already known. If knowledge that $\yrv = y$ makes $\xrv = x$ less surprising, then $\frac{1}{p(x \mid y)}$ will decrease, reflecting that $y$ tells us something about the value of $x$. For example, if all $\bigstar$'s are $\texttt{\color{blue}{blue}}$, then learning the object is a $\bigstar$ eliminates all uncertainty about its color: $p(\texttt{\color{blue}{blue}} \mid \bigstar) = 1$, so all of the information in $\xrv = \texttt{\color{blue}{blue}}$ is shared by $\yrv = \bigstar$.

The \textbf{mutual information} is the probability-weighted average of point-wise mutual information over all outcomes. It represents how much one random event's outcome typically tells us about another:

\begin{align}
    I(\xrv; \yrv) = \sum_{x \in \mathcal{X}} \sum_{y \in \mathcal{Y}} p(x, y)\log \left( \frac{p(x, y)}{p(x)p(y)} \right) 
\end{align}

Just as in \textbf{Figure \ref{what_is_information}}, where one bit of information meant ruling out half of the probability mass, on average, each bit of mutual information between $\xrv$ and $\yrv$ lets us eliminate half of the probability mass of one event after observing the other. This is most easily seen when all outcomes are equally probable, because then eliminating half the probability mass corresponds to eliminating half the outcomes.

\begin{figure*}[htbp]
\centering
\includegraphics[width=1\linewidth]{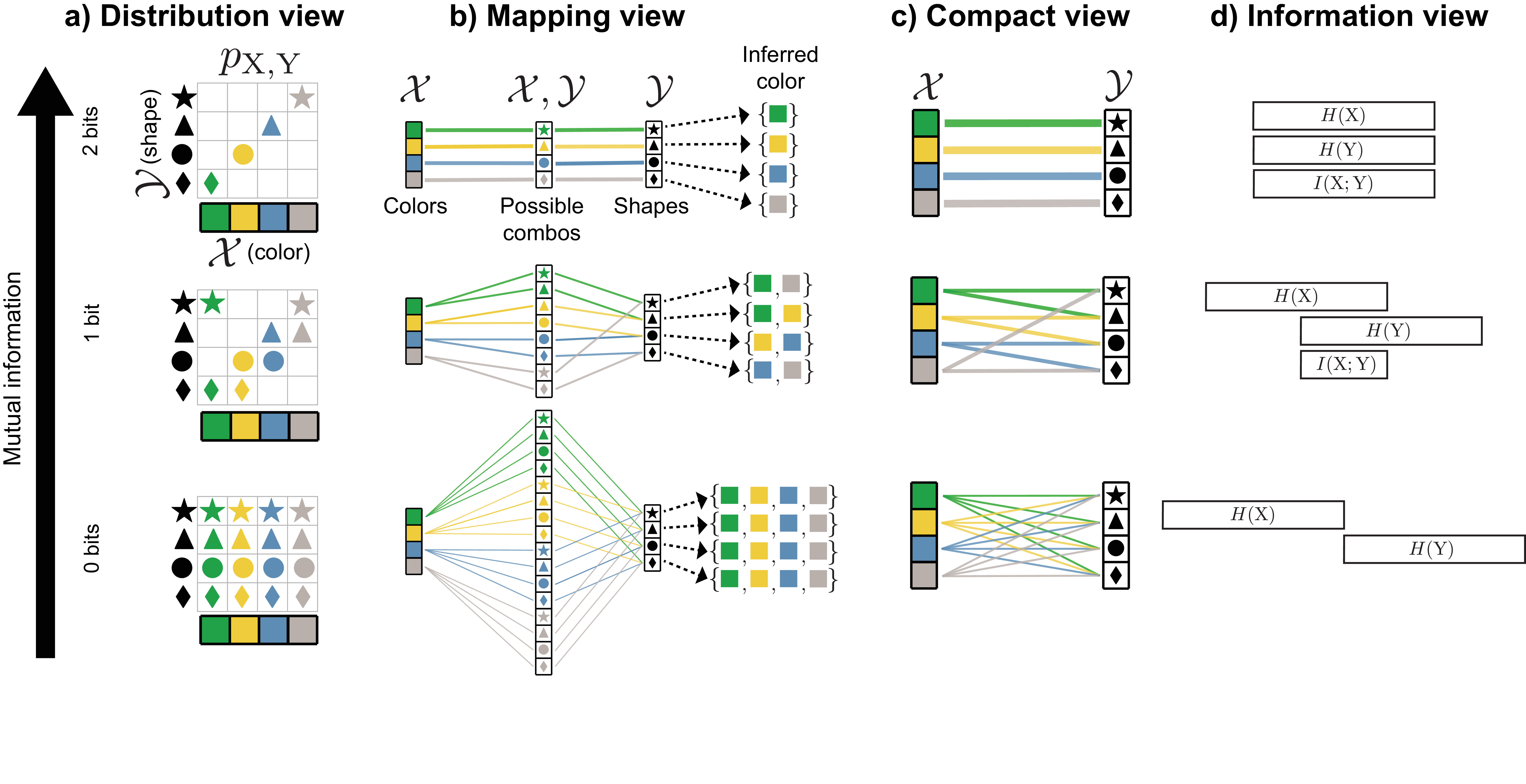}
\caption{\textbf{Mutual information} describes the relationship between two random variables. Here those random variables are the shape and color of an object drawn at random. The joint distribution of shape and color determines the amount of mutual information. (\textbf{Top row}) 2 bits of mutual information, (\textbf{middle row}) 1 bit of mutual information, (\textbf{bottom}) 0 bits of mutual information. \textbf{a)} The joint distribution of shape and color, with uniform probability over all possible shape/color combinations shown. \textbf{b)} Mapping view showing the colors, possible shape color combinations, possible shapes, and the possibilities for colors that can be inferred from shape alone. Line thickness shows strength of the relationship. \textbf{c)} Compact view that omits the joint distribution and color inference. \textbf{d)} More of the entropy of the two events is shared with greater mutual information.}
\label{what_is_mi}
\end{figure*}

\textbf{Figure \ref{what_is_mi}} illustrates three scenarios with varying amounts of mutual information. When $I(\xrv; \yrv) = 0$ bits, observing the object's shape does not allow us to eliminate any possible colors. When $I(\xrv; \yrv) = 1$ bit, we can eliminate two of the four possible colors. When $I(\xrv; \yrv) = 2$ bits, we can eliminate 3 of the four possible colors and know exactly what color it is, having gained all 2 bits of information in $\xrv$.

Since we can at most gain 2 bits from observing 1 of 4 possible shapes (i.e. $\maxentropy(\mathcal{Y}) = 2$), if there were more than 2 bits of information in $\xrv$, we wouldn't be able to determine color exactly. For example, if there were 8 possible colors that were all equally probable, we could not uniquely identify all possibilities based on 4 possible shapes.

\subsection{Joint and conditional entropy}

Two related quantities complete the picture: \textbf{conditional entropy}, which measures how much uncertainty remains about one event after observing another, and \textbf{joint entropy}, which measures the total uncertainty when describing two events together. These are essential for reasoning about information transmission, where we need to know not just how much information a signal carries, but how much survives after passing through a noisy process.

\subsubsection{Conditional entropy}

While mutual information measures how much observing $\yrv$ reduces uncertainty about $\xrv$, conditional entropy measures how much uncertainty about $\xrv$ remains after observing $\yrv$. The two sum to $H(\xrv)$, the total uncertainty in $\xrv$ alone.

As before, we start with the point-wise case. The information gained by learning the outcome of $\xrv$ after already knowing $\yrv=y$ is:

\begin{align*}
   \log \frac{1}{p(x \mid y)}
\end{align*}

Or, in the specific case of learning an object is $\texttt{\color{OliveGreen}{green}}$ after knowing it is a $\bigbullet$:

\begin{align*}
   \log \frac{1}{p({\color{OliveGreen}{\texttt{green}}} \mid \bigbullet)}
\end{align*}

The surprise of finding a $\texttt{\color{OliveGreen}{green}}$ $\color{OliveGreen}{\bigbullet}$ depends on how many different colors of $\bigbullet$ are in our set of objects. Just as we calculated entropy as a probability-weighted average over all outcomes, we can compute the average uncertainty of the object's color given that the object is a $\bigbullet$ \figrefsub{pointwise_conditional_entropy_fig}{a}. This yields the point-wise conditional entropy:

\begin{figure*}[htbp]
\centering
\includegraphics[width=\linewidth]{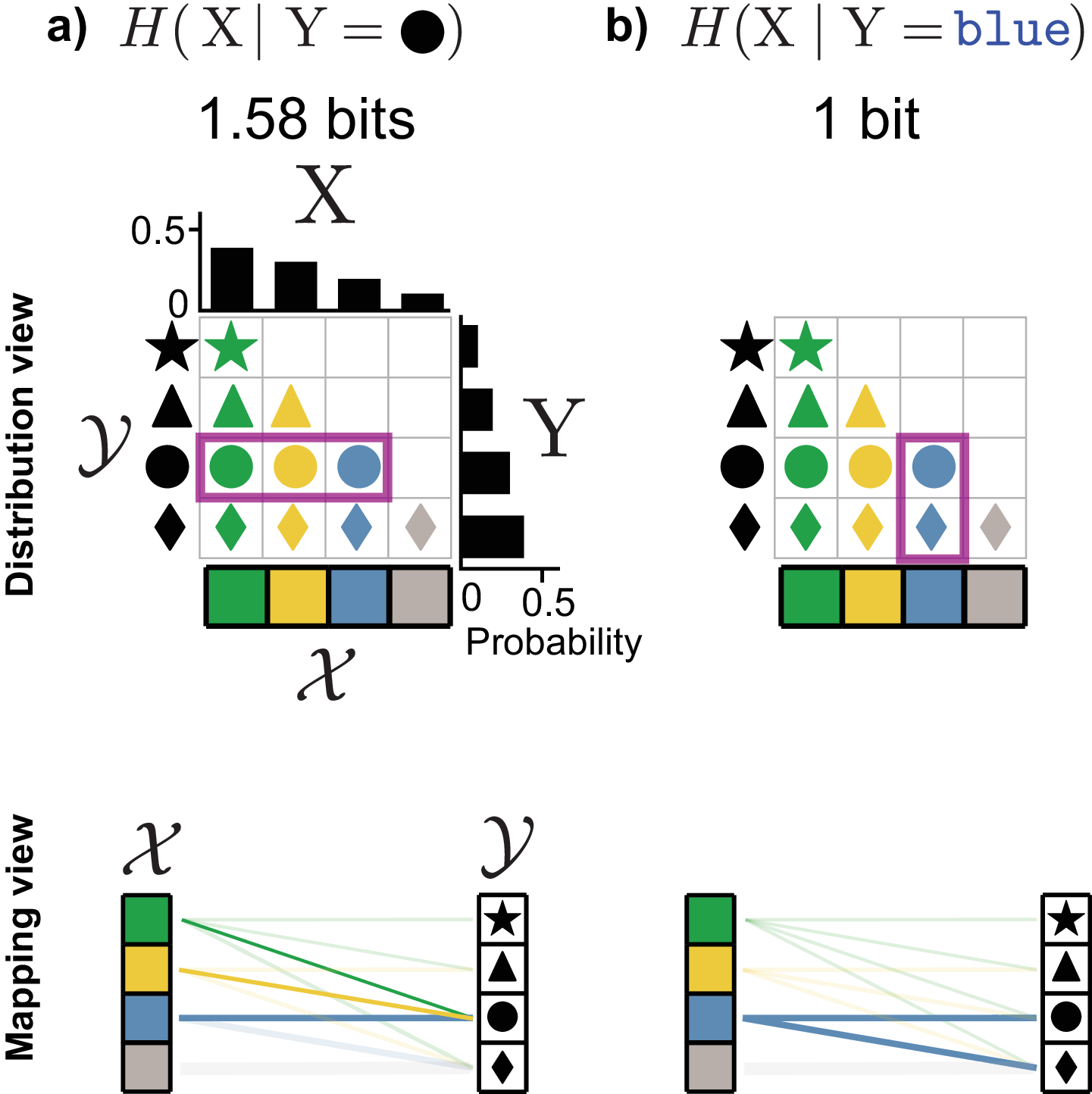}
\caption{\textbf{Point-wise conditional entropy} (\textbf{Top}) The joint and marginal distributions of shape and color. (\textbf{Bottom}) The compact mapping view between shape and color. \textbf{a)} The conditional entropy of color given the shape is $\bigbullet$ is $\log_2 3 \approx 1.58$ bits since there are three equally probable possibilities (in magenta box on distribution view) \textbf{b)} The conditional entropy of shape given a $\texttt{\color{blue}{blue}}$ object is 1 bit since there are two equally probable shapes.}
\label{pointwise_conditional_entropy_fig}
\end{figure*}

\begin{align}
   \label{conditional_entropy_point}
   H(\xrv \mid \bigbullet) = \sum_{\substack{x \in \{\texttt{\color{OliveGreen}{green}}, 
   \texttt{\color{blue}{blue}} , \\
   \texttt{\color{Dandelion}{yellow}},
   \texttt{\color{gray}{gray}}\}}}
   p(x \mid \bigbullet)
   \log \frac{1}{p(x \mid \bigbullet)} 
\end{align}

which can also be written in generic form as:

\begin{align*}
   H(\xrv \mid y) = \sum_{x \in \mathcal{X}}
   p(x \mid y)
   \log \frac{1}{p(x \mid y)} 
\end{align*}

Looking at the joint distribution in the middle row of \textbf{Figure \ref{what_is_mi}a}, knowing that the object is a $\bigbullet$ leaves three possible colors: \texttt{\color{OliveGreen}{green}}, \texttt{\color{Dandelion}{yellow}}, or \texttt{\color{blue}{blue}}. Since we've specified that these occur with equal probability, the point-wise conditional entropy will be $\log_2 3 \approx 1.58$ bits. 

Alternatively, if we looked at the conditional entropy of color given that the object is a $\bigstar$, the only possible color would be \texttt{\color{OliveGreen}{green}}. This means the point-wise conditional entropy is 0: we know the value of $\xrv$ exactly given $\yrv$.

Next, consider the average uncertainty of color, not conditional on a specific shape, but averaged over all possible shapes. We sum equation \eqref{conditional_entropy_point} over all states in $\mathcal{Y}$ (e.g. $\bigstar$,$\blacklozenge$,$\blacktriangle$,$\bigbullet$), weighting by the probability of each shape:

\begin{align*}
   H(\xrv \mid \yrv) &= \sum_{y \in \mathcal{Y}} p(y) \sum_{x \in \mathcal{X}} p(x \mid y) \log \frac{1}{p(x \mid y)}  
\end{align*}

With slight algebraic manipulation, this becomes the traditional equation for \textbf{conditional entropy}:

\begin{align*}
    H(\xrv \mid \yrv) =  \sum_{y \in \mathcal{Y}} \sum_{x \in \mathcal{X}} p(x, y) \log{\frac{1}{p(x \mid y)}}
\end{align*}

When $H(\xrv \mid \yrv)=0$, observing the outcome of $\yrv$ tells us exactly what the outcome of $\xrv$ is. When $H(\xrv \mid \yrv)>0$, some uncertainty about $\xrv$ remains after observing $\yrv$, at least for some outcomes of $\yrv$.

\subsubsection{Joint entropy}
Joint entropy is an extension of entropy to multiple random variables, which takes into account the dependence between them (i.e. the mutual information):

\begin{align*}
H(\xrv, \yrv) = \sum_{x \in \mathcal{X}} \sum_{y \in \mathcal{Y}} p(x, y) \log{\frac{1}{p(x, y)}}    
\end{align*}

For example, in \textbf{Figure \ref{what_is_mi}}, $p(x,y)$ represents the probability of a particular shape-color combination.

When $\xrv$ and $\yrv$ are independent, observing the value of $\yrv$ provides no information about $\xrv$. Thus, $H(\xrv, \yrv) = H(\xrv) + H(\yrv)$. This can be proven algebraically by substituting $p(x,y) = p(x)p(y)$ into the joint entropy formula, as shown in Appendix \ref{independent_entropies_proof}.

When $\xrv$ and $\yrv$ are dependent, the joint entropy will be less than the sum of the individual entropies, since there will be nonzero mutual information. For example, in the top row of \textbf{Figure \ref{what_is_mi}}, $H(\xrv) = 2$, $H(\yrv) = 2$, but $H(\xrv, \yrv) = 2$ also, since the 2 bits are mutual to both random variables.

\subsection{Relationships between entropy and mutual information}
\label{entropy_mi_relationship_sec}

\begin{figure*}[htbp]
\centering
\includegraphics[width=0.9\linewidth]{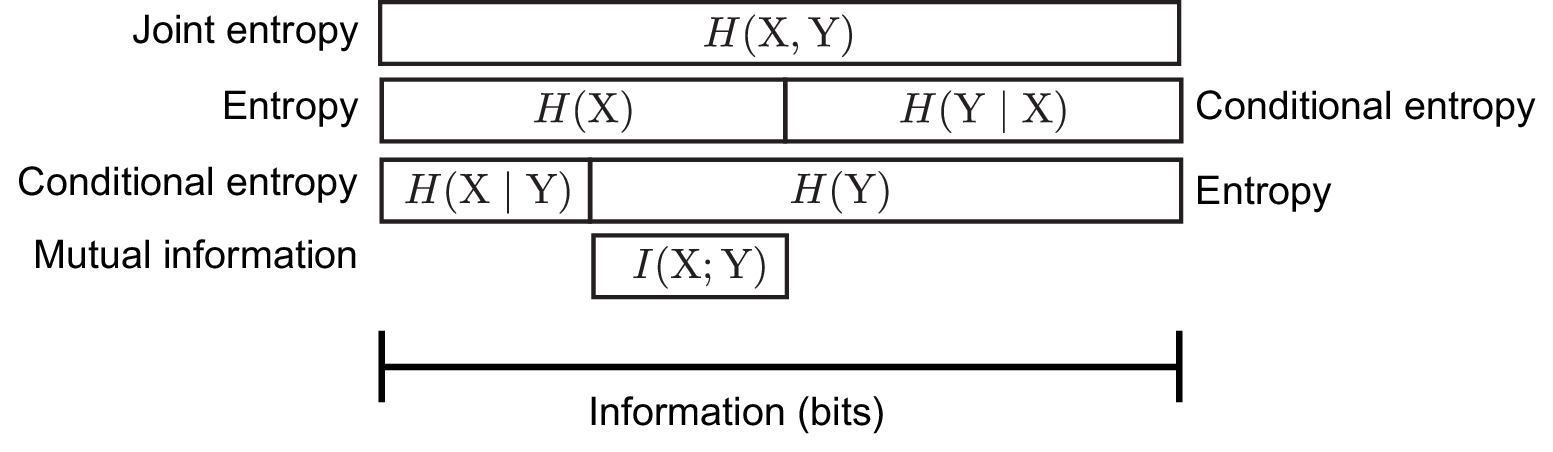}
\caption{\textbf{The relationships between entropy, joint entropy, conditional entropy and mutual information} The width of each bar represents its size (in bits).  Adapted from \cite{MacKay2005}, p140}
\label{entropy_mi_relationship}
\end{figure*}

Entropy, joint entropy, conditional entropy, and mutual information can be described in terms of each other. Their relationships are summarized in \textbf{Figure \ref{entropy_mi_relationship}}. Entropy is the average uncertainty in a single random event. Mutual information is the amount of that uncertainty in common with another random event through some statistical relationship. Conditional entropy is the remaining uncertainty in one random event after subtracting the mutual information shared with another. Joint entropy is the average uncertainty when describing both random events together.

Mutual information can be written in three different ways in terms of the other entropies:

\begin{align*}
    I(\xrv; \yrv) &= H(\yrv) - H(\yrv \mid \xrv) \\
                    &= H(\xrv) - H(\xrv \mid \yrv)  \\
                    &= H(\xrv) + H(\yrv) -  H(\xrv, \yrv) 
\end{align*}

An example of how $I(\xrv; \yrv)$ can be manipulated into these other forms can be found in Appendix \ref{mi_decomp_proof}.

The first two can be interpreted as the total uncertainty in one event minus the uncertainty that remains after observing the other. The third is the total uncertainty of both events in isolation minus the uncertainty of both considered together.

\subsection{Entropy rate of stochastic processes}
\label{entropy_rate_sec}
Thus far, we've discussed only the case in which the source of information is a sequence of independent and identically distributed (IID) random variables, but the information theoretic concepts we've introduced generalize to non-independent ordered sequences of random variables, also known as \textbf{stochastic processes}.

In the IID case, we have a random variable $\xrv$, which has some probability density/mass function $\prob{\xrv}(x)$. A sequence of events can be represented by the random vector $\bmathx = (\xrv_1, \xrv_2,...)$, which is an ordered sequence of random variables. Its joint probability density/mass function is denoted $\prob{\bmathx}(\mathbf{x}) = \prob{\xrv_1, \xrv_2,...}(x_1, x_2,...)$. The IID assumption allows us to simplify this expression by factoring the joint distribution and writing each term as the shared probability density/mass function of $\xrv$:

\begin{align*}
    \prob{\bmathx}(\mathbf{x}) &= \prob{\xrv_1, \xrv_2,...}(x_1, x_2,...) \\
            &= \prob{\xrv}(x_1)\prob{\xrv}(x_2)...
\end{align*}

Entropy is additive for independent random variables, so the entropy of $\bmathx = (\xrv_1, \xrv_2,... \xrv_N)$ will be equal to $N$ times the entropy of $\xrv$. Thus, $H(\xrv)$, the entropy per event, is called the \textbf{entropy rate} of the process. This quantity is important because a communication channel must keep up with this rate to transmit the source faithfully (Section \ref{transmission_limits}).

Alternatively, suppose that this IID assumption does not hold. What is the entropy rate of a stochastic process with an arbitrary joint probability distribution? There are two ways to define it, and under certain assumptions they are equivalent (\cite{Cover2011} Sec. 4.2).

\begin{figure*}[htbp]
\centering
\includegraphics[width=1\linewidth]{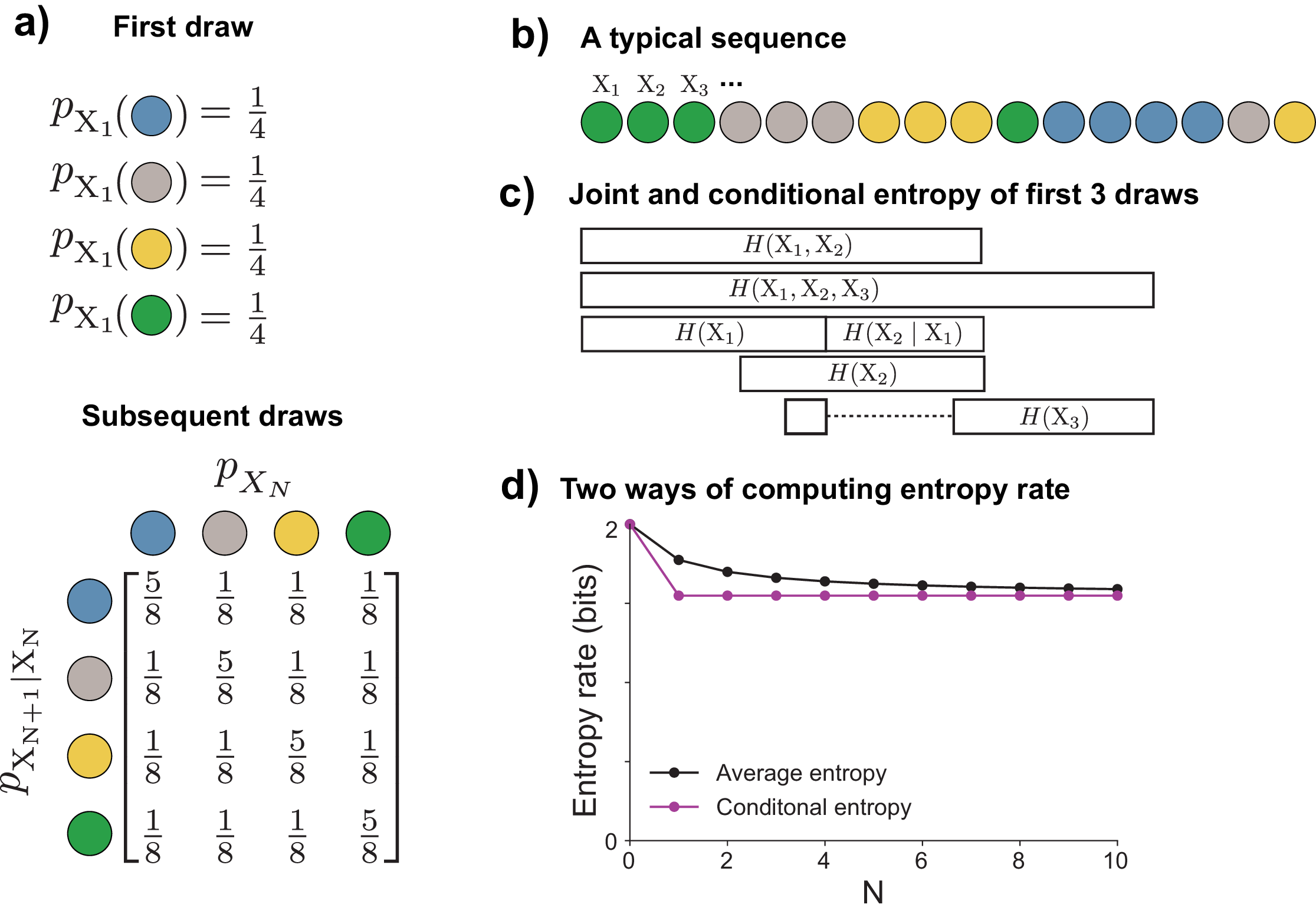}
\caption{\textbf{Entropy rate of a stochastic process} \textbf{a)} The stochastic process, which consists of \textbf{(top)} an initial draw of a colored marble with each color having a probability of $\frac{1}{4}$ and \textbf{(bottom)} subsequent draws where the probability of repeating the same color as the previous draw is $\frac{5}{8}$ and the probability of all other colors $\frac{1}{8}$. \textbf{b)} A typical sequence from this stochastic process where a random variable $\xrv_k$ represents the color selected at each position. \textbf{c)} Entropy, conditional entropy, and joint entropy of the first three draws. Knowledge of past outcomes reduces uncertainty of future outcomes (or vice versa). \textbf{d)} The two ways of computing entropy rate: the average of the joint entropy and the conditional entropy of the next draw given the previous. For a stationary stochastic process, these converge to the same value as $N \rightarrow \infty$.}
\label{entropy_rate_fig}
\end{figure*}

Returning to the example of drawing marbles, suppose we are drawing from a magical urn that, the first time we draw from it, will give all four colors with equal probability. After the first draw, the magic kicks in, and the urn makes it more likely that we will again draw the same color as our previous draw. For example, the conditional probability  $\prob{\xrv_{N + 1} \mid \xrv_{N}}(\tt{\color{blue}{blue}} \mid \tt{\color{blue}{blue}}) = \frac{5}{8}$, while $\prob{\xrv_{N + 1} \mid \xrv_{N}}(\tt{\color{OliveGreen}{green}} \mid \tt{\color{blue}{blue}}) = \frac{1}{8}$ \figrefsub{entropy_rate_fig}{a}. If we write out all the possible conditional probabilities, we will get a matrix where each column represents the conditional distribution $\prob{\xrv_{N+1} \mid \xrv_N}$. A typical sequence from this distribution will tend to give repeats of the same color \figrefsub{entropy_rate_fig}{b}.

What can we say about uncertainties in this process? We are more uncertain about the color of the first draw $\xrv_1$ (all colors with $\frac{1}{4}$ probability) than we are about the second color after observing the first ($\frac{5}{8}$ chance of the same color again). This means that $H(\xrv_1) > H(\xrv_2 \mid \xrv_1)$ \figrefsub{entropy_rate_fig}{c}. Knowing $\xrv_1$ reduces our uncertainty about the value of $\xrv_2$, and similarly knowing $\xrv_2$ would reduce our uncertainty about the value of $\xrv_1$. This means that the joint entropy is less than the sum of the individual entropies $H(\xrv_1, \xrv_2) < H(\xrv_1) + H(\xrv_2)$.

What about the \textbf{entropy rate} of the process? There are two ways we could define this. First, we could simply say that the entropy rate is our average uncertainty over all draws. This can be quantified by dividing the joint entropy by the number of draws to get an average of the information in each draw:

\begin{align}
    \frac{1}{N} H(\xrv_1, \xrv_2,...\xrv_N) \label{entropy_rate_average}   
\end{align}

Alternatively, we could say that the entropy rate is our uncertainty about the next draw, given the previous ones. This is quantified with the conditional entropy:

\begin{align}
     H(\xrv_{N+1} \mid \xrv_{N}, \xrv_{N-1},... \xrv_{1}) \label{entropy_rate_cond}
\end{align}

Under our example, \eqref{entropy_rate_average} will gradually decrease as $N$ increases, as our high uncertainty about the first draw has a proportionally smaller and smaller effect compared to the subsequent, more predictable draws. 

For the second definition, since our model specifies that draw $N+1$ only depends on the outcome of draw $N$ (a type of model called a \textbf{Markov chain}), \eqref{entropy_rate_cond} will simplify to:

\begin{align}
     H(\xrv_{N+1} \mid \xrv_{N})
\end{align}

Thus, the unconditional entropy of the first draw $H(\xrv_1) = 2$ bits is large, while the conditional entropy $H(\xrv_{N+1} \mid \xrv_N)$ is lower and constant for all subsequent draws.

As $N \rightarrow \infty$, both measures will converge to the same value for the entropy rate because the high uncertainty of the first draw is ultimately neglected for both, so the two definitions are equivalent in the limit. This will be true whenever the stochastic process is \textbf{stationary}, which means that its joint probability distribution is shift-invariant. Mathematically, a stationary process has the property:

\begin{align*}
    \prob{\xrv_1, \xrv_2, ..., \xrv_N} = \prob{\xrv_{1+k}, \xrv_{2+k}, ..., \xrv_{N+k}}
\end{align*}

for any positive integer $k$.

The process in \textbf{Figure \ref{entropy_rate_fig}} is stationary: the uniform initial distribution is the stationary distribution of this symmetric chain, so the joint distribution is shift-invariant.

\subsubsection{Redundancy and stochastic processes}

Moving from a sequence of independent and identically distributed random variables to a stochastic process requires revisiting the definition of \textbf{redundancy} provided in section \ref{redundancy_sec}. We can no longer define redundancy with respect to only the entropy of a single random variable in the sequence (e.g. $H(\xrv_k)$). The more general definition of redundancy must account for the joint entropy:

\begin{align*}
    W(\bmathx) = \maxentropy(\mathcal{X}\times\mathcal{X}\times...) - H(\xrv_1, \xrv_2,...)
\end{align*}

Redundancy now depends on higher-order interactions between variables, not just on each variable's individual distribution. In the example above, $H(\xrv_k) = \maxentropy(\mathcal{X})$: all colors are equally likely for every draw in the absence of any information about the outcomes of other draws. But the joint distribution is not the maximum entropy distribution, because the information present in one random variable is shared by others. This is why we can do much better than random chance in guessing the next color given knowledge of the previous ones. The maximum entropy joint distribution would be independent and identically distributed random events, each with equal probability over all states.

\subsection{Probability densities and differential entropy}
\label{differential_entropy_sec}

The examples above all considered probability mass functions over finite, discrete sets of events. How do the formulae and interpretations generalize to cases other than this, such as probability density functions defined over the infinite real line?

For probability densities that are defined over the continuous real line (e.g. the normal distribution, the exponential distribution), we can define an analog to entropy called the \textbf{differential entropy}, replacing the discrete sum with an integral:

\begin{align}
    H(\mathrm{X}) = -\int_{\mathcal{X}}p(x)\log p(x)dx \label{diff_entropy}
\end{align} 

where $p(x)$ is the probability density function of X.  

The differential entropy does not have the same clear interpretations as the discrete entropy. In particular, it cannot be interpreted as the number of bits needed to describe an outcome, since specifying a real number with arbitrary precision requires infinitely many bits. The differential entropy can also take negative values unlike its discrete counterpart. Finally, as pointed out in (\cite{MacKay2005} p180), equation \eqref{diff_entropy} is improper in that it takes the logarithm of a dimensional quantity (the probability density has units $\frac{1}{x}$), so the value of differential entropy changes when units change.

One way to resolve these difficulties is to discretize the probability density by dividing the real line into equally spaced intervals of width $\Delta$ and integrating within each interval to obtain a probability mass function (\cite{Cover2011} section 8.3). One can then compute the discrete entropy of this distribution, which will be equal to the differential entropy minus $\log \Delta$. In other words, the more fine grained our discretization of the probability density, the more bits are needed to describe it. In many cases we may work directly with such discretized probability densities on a computer. The absolute value of the entropy will then differ from the differential entropy by an additive constant determined by the discretization width.

The maximum entropy distribution for a given state space also works differently for continuous probability densities. If our state space $\mathcal{X}$ is the set of real numbers $\mathbb{R}$, then the entropy can be driven infinitely high by making an infinitely wide distribution. However, we can still describe a maximum entropy distribution under certain constraints. For example, the maximum entropy distribution on $\mathbb{R}$ with a variance equal to $\sigma^2$ is a normal distribution with that variance. Many commonly used distributions are optimal under similar constraints, and maximum entropy distributions have many important applications in statistical inference and statistical physics that are beyond our scope here.

Fortunately, mutual information between continuous probability densities does not suffer from the same difficulties. Even though it may take an infinite number of bits to specify an arbitrary real number, the interpretation that $1$ bit of information allows you to rule out half of the probability mass remains valid. Mathematically, mutual information for probability densities is equivalent to the discrete case, with sums replaced by integrals:

\begin{align*}
    I(\xrv; \yrv) = \int_{\mathcal{X}} \int_{\mathcal{Y}} p(x, y)\log \left( \frac{p(x, y)}{p(x)p(y)} \right) dxdy
\end{align*}

Mutual information also avoids the improper logarithm problem, because the log is taken on a ratio of densities in which the units cancel. Mutual information can also be computed between discrete and continuous random variables \cite{Ross2014}.

In summary, despite the difficulties of differential entropy, mutual information retains a clear interpretation as the amount of information one random variable provides about another.

\subsection{Rate distortion theory}
\label{rate_distortion_sec}
What happens if we try to compress data to a size smaller than its information content, or infer the outcome of a random event about which we have incomplete information? These questions can be answered by \textbf{rate-distortion theory}.

One application of rate distortion theory is lossy compression, where we compress a source to the smallest size possible while tolerating some error upon decompression. That is, the decompressed message will be ``close to" but not exactly the same as the compressed source. Because we can tolerate some error, we need not preserve all of the information about the source in the message. 

\begin{figure*}[htbp]
\centering
\includegraphics[width=0.8\linewidth]{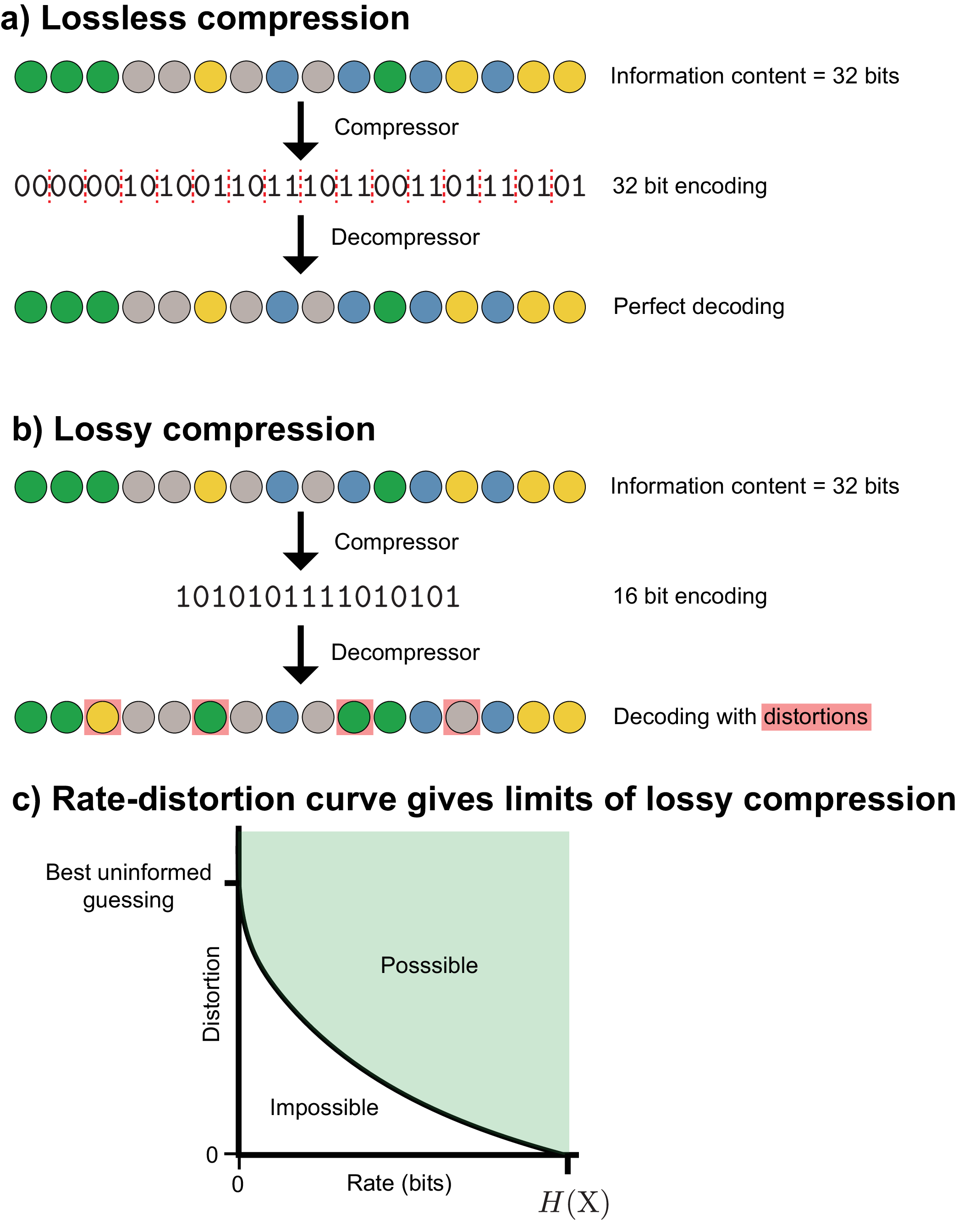}
\caption{\textbf{Lossy compression and rate distortion} \textbf{a}) In lossless compression, a typical sequence is mapped to a binary codeword with length equal to its information content and can be decompressed without error. \textbf{b}) In lossy compression, a sequence is mapped to a sequence shorter than its information content, and errors are present upon decompression. \textbf{c}) The black curve shows the minimum number of bits needed to achieve a given average distortion.}
\label{rate_distortion_fig}
\end{figure*}

\textbf{Figure \ref{rate_distortion_fig}} provides an example of this. We have some source of information, like the equally probable (maximum entropy) four-color marble scenario of \textbf{Figure \ref{what_is_entropy}}. A compressor maps a series of random colors to the shortest binary string possible (on average), and a decompressor tries to reproduce the original sequence from that string. In \textbf{lossless compression} \figrefsub{rate_distortion_fig}{a}, we impose the requirement that the decompressed binary string must match exactly the original sequence. In \textbf{lossy compression} \figrefsub{rate_distortion_fig}{b}, we're willing to tolerate some errors upon decompression, which will allow us to discard some of the source information and achieve an even shorter binary encoding.

There is a trade-off between how much information we discard and how many errors will be present after decompression: the more information we throw away, the shorter a binary encoding we can make, but the more errors we will make upon decompression. The \textbf{rate-distortion curve} quantifies this trade-off. The rate quantifies the amount of information we have about the source, and distortion quantifies the number and severity of the errors we make. The rate-distortion curve applies not just to data compression problems, but to any situation in which we have incomplete information about a random event, such as the imperfect transmission of data.

A \textbf{distortion} function $d(x, \hat{x})$ measures how much an original event $x \in \mathcal{X}$ (e.g. a color) differs from its decompressed version $\hat{x} \in \mathcal{\hat{X}}$, returning a non-negative real number:

\begin{align*}
    d : \mathcal{X} \times \mathcal{\hat{X}} \rightarrow \mathbb{R}^+
\end{align*}

In the example shown in \textbf{Figure \ref{rate_distortion_fig}}, $\mathcal{X}$ and $\mathcal{\hat{X}}$ are identical (i.e. the set of the four possible colors), but this may not be true in every situation. For example, we might consider the distortion associated with storing an arbitrary real number as a 32-bit floating point number on a computer. In this case, $\mathcal{X}$ would be the set of real numbers and $\mathcal{\hat{X}}$ would be the set of possible 32-bit floating point numbers.

The source produces information at some rate, which we'll call the ``source rate," whereas the \textbf{rate} describes how much of it we capture (over, for example, time, space, etc.). Higher rates are better, because less source information is lost. For a discrete source, the rate cannot exceed the source's entropy $H(\xrv)$, since no more information exists to capture.

A useful summary of the distortion is the expected value of the distortion function over the source. For example, this could be the average number of differences between an original sequence and the sequence obtained by passing it through a lossy encoder and decoder (as shown in \figrefsub{rate_distortion_fig}{b}):

\begin{align*}
	\mathrm{D} = \e{d(\xrv, \hat{\xrv})}
\end{align*}

With complete information, distortion can be zero; with zero information, zero distortion is impossible (unless the source is constant). More information allows lower distortion. Rate-distortion theory makes this precise: to achieve distortion below some threshold $\mathrm{D}_{\text{max}}$, we must have at least $R(\mathrm{D})$ bits of information about $\xrv$, where $R(\mathrm{D})$ is the \textbf{rate-distortion function}. If we are already achieving the best average distortion for the amount of information we have, $I(\xrv; \hat{\xrv})$, we cannot do better unless we acquire more information. This holds regardless of the chosen distortion function, provided that $d(x, \hat{x})$ is finite for every possible $x$ and $\hat{x}$, and is proven in the asymptotic case of infinitely long block lengths.

This doesn't necessarily mean that having  $N$ bits about a source will give us an $\hat{\xrv}$ that achieves the $N$-bit best performance for a given distortion function. For example, if we know $\xrv$ perfectly, and we applied an invertible operation like mapping each color to a new, distinct color, we would not have lost any information. However, if our distortion function was the number of incorrect colors, the average distortion would increase.

An example rate-distortion function for a discrete source is shown in \textbf{Figure \ref{rate_distortion_fig}c}. The green area represents possible combinations of rates and distortions. For sufficiently high values of average distortion, no information about the source is necessary, so the curve intersects the horizontal axis. For discrete sources, when $d(x, \hat{x}) = 0$ if and only if $x = \hat{x}$, the curve intersects the vertical axis at the value of the source's entropy, at this point we have zero remaining uncertainty. In contrast, for a continuous source, the curve would asymptotically approach the vertical axis, since an infinite number of bits would be needed to describe a continuous source exactly (Sec. \ref{differential_entropy_sec}).

In general, $R(\mathrm{D})$ will be a non-increasing convex function of $\mathrm{D}$, which means that 1) achieving lower and lower distortion will always require acquiring more information (assuming we are already achieving the best possible distortion for our current rate), and 2) there will be diminishing returns on lowering distortion as we acquire more and more information.

\section{Channels}

A \textbf{channel} is a mapping from inputs in $\mathcal{X}$ to outputs in $\mathcal{Y}$, defined by the conditional probability distribution $\prob{\yrv \mid \xrv}$. Rather than using the term ``outcome" to describe the value taken by a random variable as in the previous section, we now use the terms ``input" and ``output". When $\prob{\yrv \mid \xrv}$ can be described by a deterministic function, we have a \textbf{noiseless channel}; otherwise, we have the more general case of a \textbf{noisy channel}.\footnote{This is true of a ``memoryless'' channel, which means that multiple channel uses are independent and identically distributed. All channels discussed here assume memorylessness}

The concepts developed in the previous section (entropy, mutual information, conditional entropy) naturally lead to the question: what happens when information must travel through an imperfect medium? Channels formalize this scenario. Often, they model a physical medium like a band of radio frequencies or transistors on a computer.

\subsection{Channels as matrices}

\begin{figure*}[htbp]
\centering
\includegraphics[width=0.7\linewidth]{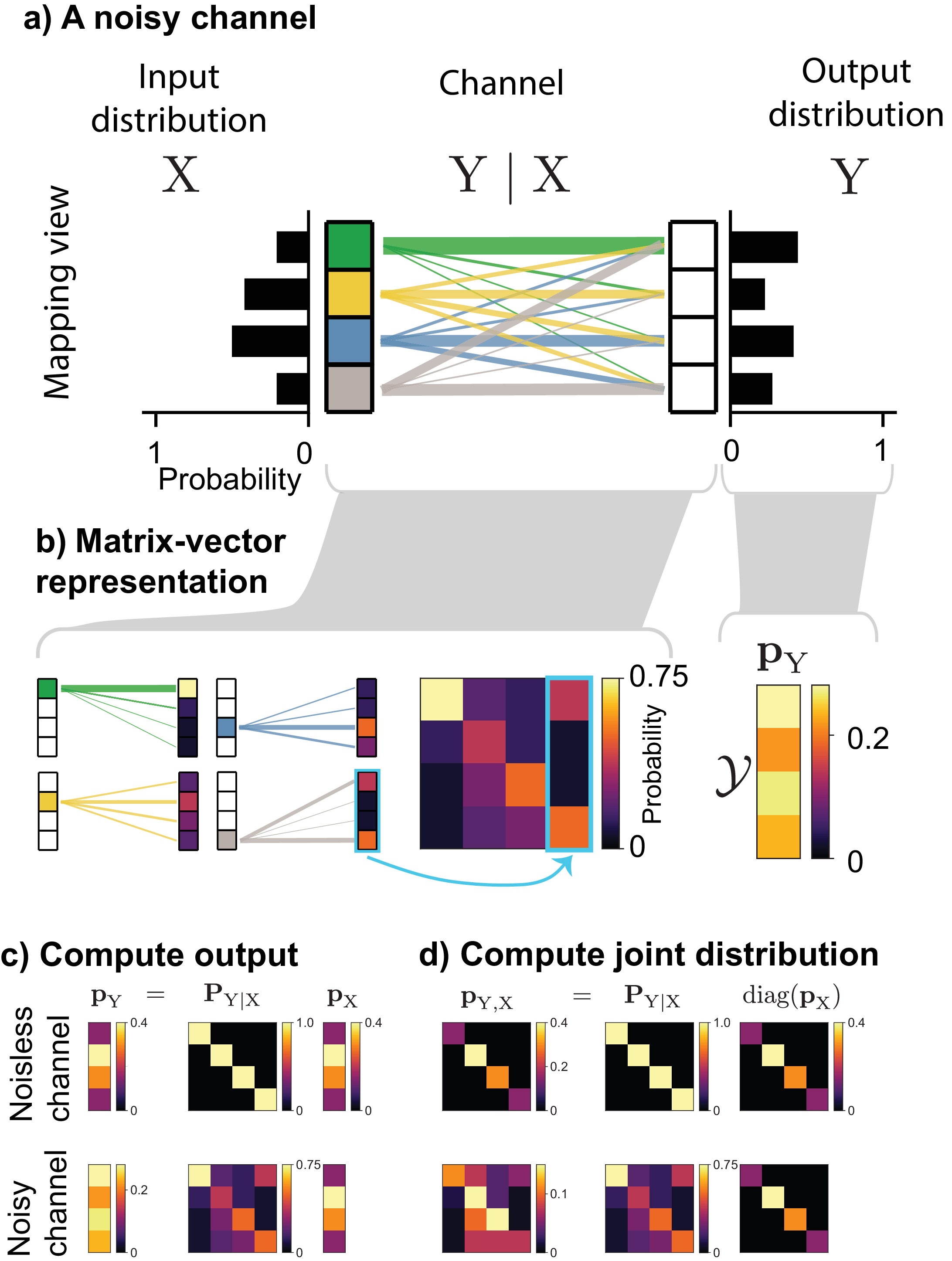}
\caption{\textbf{Channels}  \textbf{a)} The mapping view of a noisy channel. The channel is mathematically represented by the conditional random variable $\yrv \mid \xrv$ \textbf{b)} Channels and inputs/outputs can also be represented by matrices and vectors of probabilities, respectively. The vectors sum to one since they represent probability distributions, and each column of the matrix sums to one since it represents the conditional distribution $\yrv \mid \xrv=x$. The matrix/vector representations can be used to compute \textbf{c)} the output distribution and \textbf{d)} the joint distribution through matrix-vector and matrix-matrix multiplications, respectively. \textbf{d)} The joint distribution is computed from the matrix-matrix product of the channel and a diagonal matrix of the input probabilities. }
\label{what_are_channels}
\end{figure*}

A noisy channel is equivalent to a conditional probability distribution $\prob{\yrv \mid \xrv}$, and for a discrete outcome space, this can be visualized either as a mapping or a matrix \figrefsub{what_are_channels}{a}. In the matrix form, which we denote $\mathbf{P}_{\yrv \mid  \xrv}$, each column is the conditional distribution $p(y \mid \xrv = x)$, which describes the probability that the input $x$ will map to each possible output in $\mathcal{Y}$. This describes the noise distribution of that input.  Since each column is itself a probability distribution, it sums to $1$.

The channel matrix lets us compute two quantities of interest. The output distribution $\mathbf{p}_{\mathrm{Y}}$ is the product of the channel matrix and the input distribution $\mathbf{p}_{\mathrm{X}}$ \figrefsub{what_are_channels}{c}:

\begin{align*}
    \mathbf{p}_\mathrm{Y} = \mathbf{P}_{\yrv \mid  \xrv}\mathbf{p}_\mathrm{X}
\end{align*}

The joint distribution $\mathbf{P}_{\xrv,  \yrv}$, needed for computing conditional entropy and mutual information, can be found by multiplying the channel matrix by a diagonal matrix of the input probabilities \figrefsub{what_are_channels}{d}:

\begin{align*}
    \mathbf{P}_{\xrv,  \yrv} = \mathbf{P}_{\yrv \mid  \xrv}\text{diag}(\mathbf{p}_\mathrm{X})
\end{align*}

Unlike $\mathbf{P}_{\yrv \mid \xrv}$, for which each column is a probability distribution and sums to $1$, all entries of $\mathbf{P}_{\xrv, \yrv}$ together sum to $1$ since the whole matrix is a joint probability distribution.

\subsection{Entropy and mutual information in channels}
\label{ent_and_mi_in_channels}

In a channel, entropy, conditional entropy, and mutual information take on more specific roles: they distinguish signal from noise and quantify how much information survives transmission. Our goal is to recover as much of $H(\xrv)$, the average information at the channel inputs, as possible. The mutual information between $\xrv$ and $\yrv$ measures the information shared between a transmitted and received message. The higher it is, the more information has successfully been transmitted to the receiver.

A key distinction in channels is between information that helps us infer the state of the source and randomness that is irrelevant (both are measured in bits, since the definition of entropy is based on probabilities alone). For a noisy channel, $\prob{\xrv}$ contains the information we're interested in, and $\prob{\yrv \mid \xrv}$ is the noise imparted by the channel. Both $H(\xrv)$ (the input) and $H(\yrv \mid \xrv)$ (the noise) qualify as information in the mathematical sense, even though only the former is useful. The entropy of the channel output $H(\yrv)$ will usually contain both information about the input and noise.

The information transmitted through a channel depends on both the channel itself ($\prob{\yrv \mid \xrv}$) and the distribution over the channel's inputs ($\prob{\xrv}$). Two factors govern how much information is transmitted through the channel: the noise level at the output of individual channel inputs, and how much the outputs of different inputs overlap.

The noise level at each input's output can be computed from knowledge of the channel alone without knowledge of the input distribution. For input $x$, the noise at the output is $H(\yrv \mid x)$. Assuming a particular input distribution allows us to compute the average noise imparted by the channel $H(\yrv \mid \xrv)$. Input distributions that put more probability on inputs with less noisy outputs will tend to transmit more information.

The other important factor is how much different inputs overlap at the channel output. Unlike the noise level of a particular input, overlap can only be calculated with respect to a particular input distribution, because it requires the joint distribution of inputs and outputs. The average uncertainty about which input produced a given output is $H(\xrv \mid \yrv)$. Channels with less overlap at the output transmit more information.

\paragraph{Examples of noisy and non-noisy channels}

\begin{figure*}[htbp]
\centering
\includegraphics[width=1\linewidth]{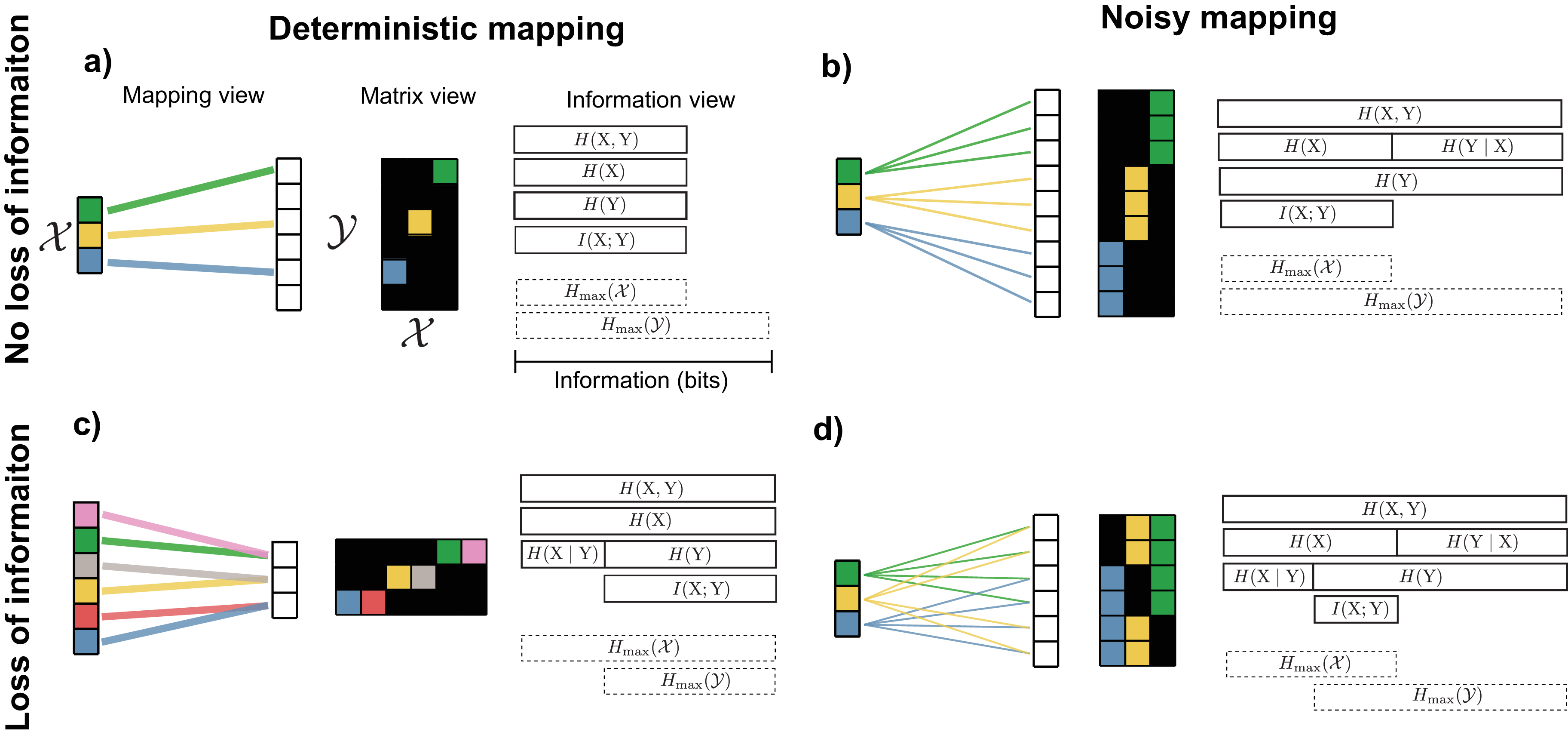}
\caption{\textbf{Examples of noisy and noiseless channels.} A random variable $\xrv$ with equal probability over all of its states is mapped to a random variable $\yrv$. These mappings can be either deterministic or noisy and information-preserving or information-destroying. Bars to the right of each mapping show entropy, conditional entropy, joint entropy, mutual information, and the maximum entropy over the state space of each random variable. Line thickness denotes magnitude of $p(y \mid x)$ \textbf{a)} A deterministic, information-preserving mapping with more outputs than inputs. Each input maps uniquely to exactly one output. The outputs with no line connected have zero probability. \textbf{b)} An information-preserving, noisy mapping. Each input maps into multiple outputs, but the outputs mapped to by a particular input are all disjoint. \textbf{c)} A deterministic, information-destroying mapping. Multiple inputs collide on the same output, meaning the mapping cannot be inverted. \textbf{d)} A noisy, information-destroying mapping. Each input maps to multiple outputs, and the outputs mapped to by each input are not disjoint.}
\label{noisy_noiseless_mappings}
\end{figure*}

We now give specific examples of channels that are either noisy or noiseless and either lose information about the inputs or preserve all of it. For simplicity, here we assume distribution over inputs is uniform. The ideal information transmission system would lose no information and be able to recover $\xrv$ exactly given an output $\yrv$, which would mean that $I(\xrv;\yrv) = H(\xrv)$. One way this can occur is in a noiseless channel in which the channel maps every input to a unique output \figrefsub{noisy_noiseless_mappings}{a}. In this situation, $H(\xrv) = H(\yrv) = I(\xrv; \yrv)$ and $H(\yrv \mid \xrv) = H(\xrv \mid \yrv) = 0$, meaning that we could perfectly predict $\xrv$ given $\yrv$ (or $\yrv$ given $\xrv$).

However, a channel does not need to be noiseless to completely preserve input information. All input information can be recovered through a noisy channel, so long as each input maps to outputs that don't overlap with those of other inputs \figrefsub{noisy_noiseless_mappings}{b}. In this scenario, $H(\yrv \mid \xrv)$ is no longer zero, because even knowing the input, we cannot predict exactly which output the noise will produce. In contrast, $H(\xrv \mid \yrv) = 0$ because given the output, we can know exactly what the input was. $H(\yrv)$ is greater than $H(\xrv)$ because it includes all of the entropy $H(\xrv)$ along with some additional entropy caused by noise.

Alternatively, even if the channel imparts no noise, information can be lost. This happens if multiple inputs map to the same output \figrefsub{noisy_noiseless_mappings}{c}. It thus becomes impossible to tell with certainty from the output $\yrv$ which input was used, or equivalently, $H(\xrv \mid \yrv) > 0$.

Finally, the most common situation in practice involves both a loss of information ($H(\xrv \mid \yrv) > 0$ or equivalently $H(\xrv) > I(\xrv; \yrv)$) and noise present in the output ($H(\yrv \mid \xrv) > 0$ or $H(\yrv) > I(\xrv; \yrv)$) \figrefsub{noisy_noiseless_mappings}{d}. Although \textbf{Figure \ref{noisy_noiseless_mappings}d} shows a scenario where $H(\yrv) > H(\xrv)$, this can also occur even if $H(\xrv) = H(\yrv)$: some of the input entropy is replaced by noise.

To summarize: $I(\xrv; \yrv)$ is the average amount of information successfully transmitted through the channel. $H(\yrv \mid \xrv)$ is the average noise at the output: the uncertainty about what the output outcome will be knowing the specific input used. $H(\xrv \mid \yrv)$ is uncertainty in which input was used knowing the output outcome, which can arise from deterministic many-to-one mappings in the channel or inputs whose noise overlaps at the output.

The different ways of decomposing mutual information presented in Section \ref{entropy_mi_relationship_sec} have more specific interpretations in the context of a noisy channel:

\begin{align*}
    I(\xrv; \yrv) = H(\xrv) - H(\xrv \mid \yrv)
\end{align*}

can be interpreted as the average input information minus the uncertainty at the output about which message was sent, determined by how much different messages overlap at the channel's output.

\begin{align*}
    I(\xrv; \yrv) = H(\yrv) - H(\yrv \mid \xrv)
\end{align*}

can be interpreted as the total uncertainty at the output (the sum of input uncertainty and noise), minus the noise in the received messages.

\subsection{Data processing inequality}
\label{data_proc_ineq_sec}

The \textbf{Data Processing Inequality} states that no physical or computational operation can increase the information about a signal. In a series of channels, information about the original input can only be preserved or lost at each step: we can only become equally or more uncertain after each channel.

Mathematically, if we have $\mathrm{A} \rightarrow \mathrm{B} \rightarrow \mathrm{C} $, where each arrow represents a channel, we say that these three variables form a \textbf{Markov chain}. The theorem states that:

\begin{align*}
    I(\mathrm{A}; \mathrm{B}) \geq I(\mathrm{A}; \mathrm{C})
\end{align*}

and this statement can be generalized to greater than three random variables.

\subsection{Maximizing information throughput}
\label{max_info_throughput_sec}

If all channel inputs 1) have equally noisy outputs and 2) have outputs that overlap with the outputs from other inputs equally, then a uniform distribution over the inputs will maximize the amount of information the channel transmits. This is because when all inputs have equally noisy outputs, $H(\yrv \mid \xrv)$ is independent of the input distribution, so maximizing $I(\xrv; \yrv) = H(\yrv) - H(\yrv \mid \xrv)$ reduces to maximizing $H(\yrv)$. When outputs also overlap equally, the symmetry ensures that a uniform input distribution achieves this maximum.

In the more general case, both of these conditions will not be met, which leads to the questions: 1) What are the optimal input distribution(s) $\mathbf{p}_{\xrv}^*$ for maximizing the channel's information throughput? 2) How much information will the channel be able to transmit if this optimal input distribution is used? The answer to the latter question is called the \textbf{channel capacity} and is denoted by $C$.  

The quantities can be found by solving an optimization problem:

\begin{align}
\label{capacity_eqn}
\mathbf{p}_{\xrv}^* = \underset{\mathcal{P}_{\mathcal{X}}}{\arg \max} \: I(\xrv; \yrv) \\
C = \underset{\mathcal{P}_{\mathcal{X}}}{\max} \: I(\xrv; \yrv) \\
\end{align}

where $\mathcal{P}_{\mathcal{X}}$ is the set of probability distributions on $\mathcal{X}$. When $\mathcal{X}$ is discrete, this is the set of nonnegative vectors whose elements sum to $1$ (formally known as the \textbf{probability simplex}). The computational details of how to solve this optimization problem can be found in Appendix \ref{optimizing_mutual_info}. Here, we focus on the intuition behind its objective function.

\begin{figure*}[htbp]
\centering
\includegraphics[width=1\linewidth]{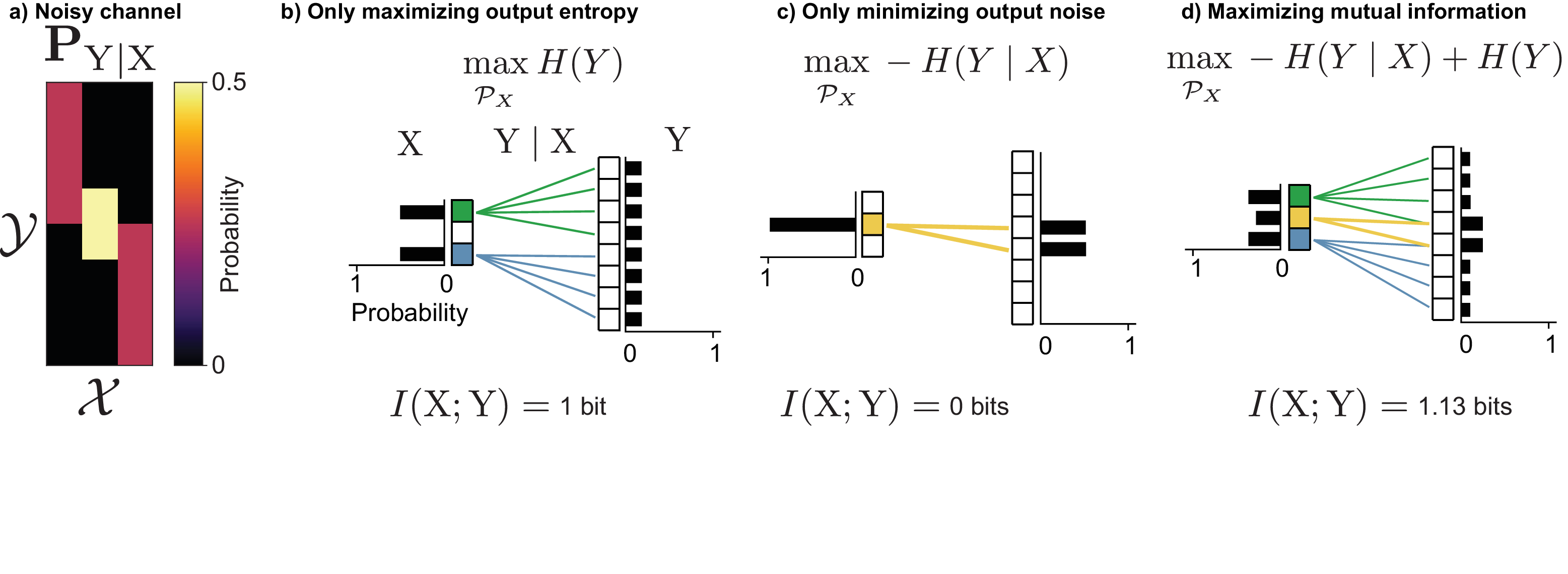}
\caption{\textbf{Optimizing mutual information for a fixed channel} gives both the optimal input distribution and the channel capacity. \textbf{a)} Matrix representation of a noisy channel. \textbf{b)} Only maximizing output entropy disregards putting probability on the input with the least noisy outputs. \textbf{c)} Only minimizing the average input noise fails to utilize the full output space. \textbf{d)} The full objective function balances these competing goals.}
\label{mutual_info_optimization_fig}
\end{figure*}

Mutual information can be decomposed as $I(\xrv; \yrv) = H(\yrv) - H(\yrv \mid \xrv)$, and this decomposition is useful in analyzing the competing goals of this optimization problem. Consider the noisy channel shown in \textbf{Figure \ref{mutual_info_optimization_fig}a}, which has two inputs that are noisy but map to distinct outputs, and one input that is less noisy but has outputs that overlap more with the outputs of the other inputs.

If we were to not maximize mutual information, but instead only maximize the first term in the decomposition, $H(\yrv)$, probability would be placed on inputs in such a way that the output probability was as uniformly distributed as possible. This is beneficial because it means more of the channel's outputs are utilized, which means different channel inputs have more space to avoid overlapping on the same outputs. For this channel, this would result in putting half the probability mass on each of the two noisy but non-overlapping inputs, which would result in $1$ bit of information gained with each use of the channel \figrefsub{mutual_info_optimization_fig}{b}.

Alternatively, if we were to minimize the noise term $H(\yrv \mid \xrv)$, probability would be placed on inputs with the least noisy outputs. Since the middle input has less noisy outputs than the other two, this would result in all of the probability mass being placed on it. With only one input ever used, no information is transmitted regardless of that input's noise level.

By optimizing the full objective, these goals are balanced: some probability is placed on the middle input (which has the least noisy outputs), while some probability is placed on the inputs with noisier outputs, which makes use of the full output space. This leads to an information transmission of $1.13$ bits, which is the channel capacity $C$.

\section{Channel coding}

In information transmission problems, there will usually be a source of random events $\mathrm{S}$. We will refer to the outcomes of $\mathrm{S}$ as \textbf{messages}. We don't have any control over the messages or the channel, but we do have the ability to design a function called an \textbf{encoder}, which will map certain messages to certain channel inputs.

The central problem of \textbf{channel coding} is: how should we design encoders to maximize information transmission and minimize the probability of incorrectly inferring the source message? The noisy channel coding theorem answers this by proving that reliable transmission is possible at any rate below channel capacity, and impossible above it. Achieving this limit requires encoding multiple messages together and carefully designing decoders to recover the original message from noisy outputs.

\subsection{Encoders}
An \textbf{encoder} is a deterministic function that maps messages from the source to the inputs of a channel. We'll make the assumption, unless otherwise stated, that every message is mapped to a unique input, so no information is lost from $\mathrm{S}$ to $\xrv$. A good encoder will map $\mathrm{S}$ to $\xrv$ so as to maximize $I(\xrv; \yrv)$, balancing the same two goals discussed in Section \ref{max_info_throughput_sec}: favoring inputs with less noisy outputs while spreading outputs to minimize overlap.

\begin{figure*}[htbp]
\centering
\includegraphics[width=0.8\linewidth]{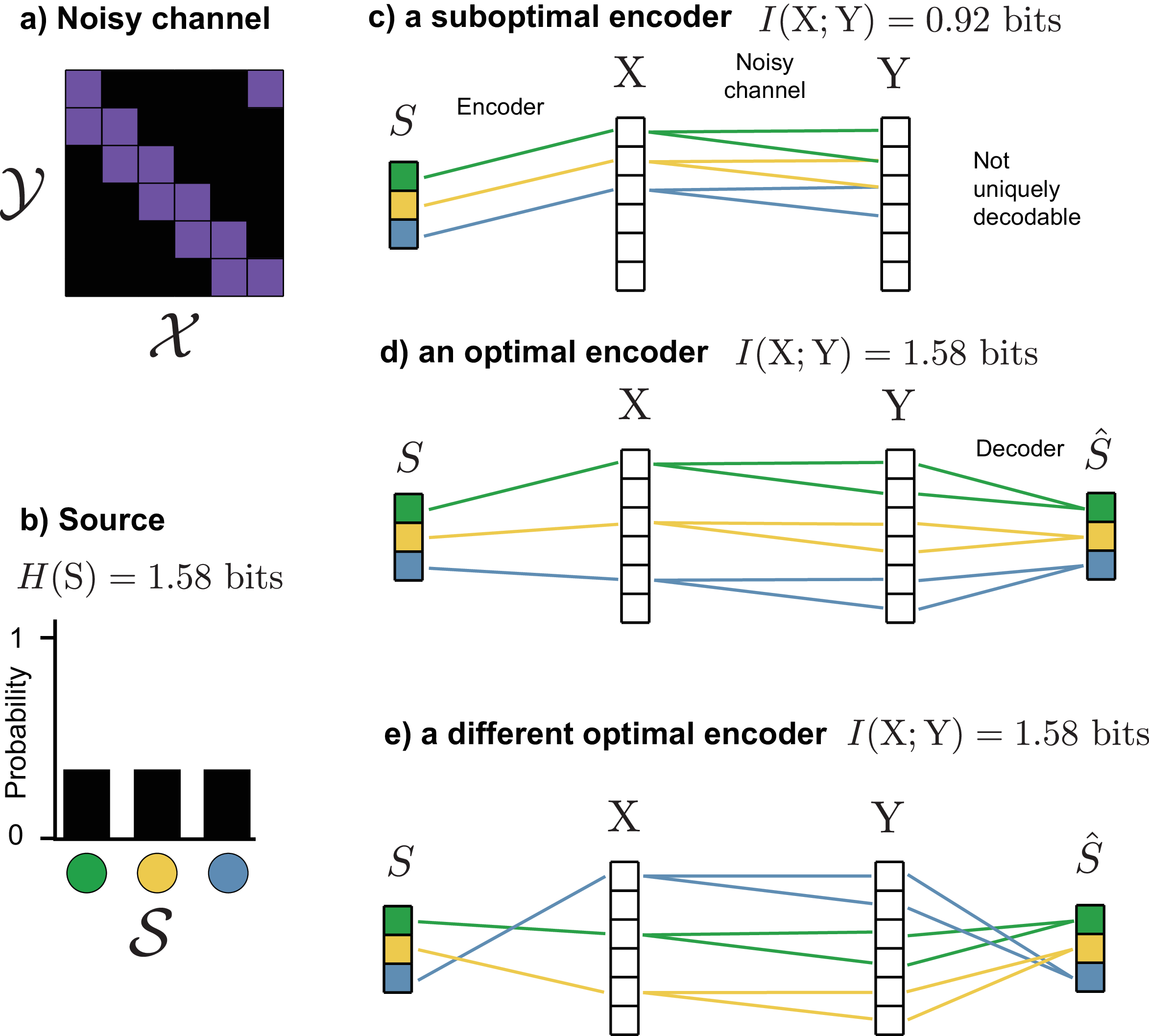}
\caption{\textbf{Optimal encoders for a noisy channel}. \textbf{a)} a noisy channel that maps each input to two possible outputs \textbf{b)} A maximum entropy source of random messages. \textbf{c)} A encoder that maps messages in $\mathcal{S}$ to inputs in $\mathcal{X}$ that produce overlapping outputs that are not uniquely decodable, thus transmitting less information than the maximum possible. \textbf{d)} An encoder that transmits the maximum amount of information by mapping messages to inputs that produce disjoint, and thus decodable, outputs. \textbf{e)} Another encoder/decoder pair that transmits the maximum amount of information, showing that the optimal encoder is not always unique.}
\label{channels_and_optimal_codes}
\end{figure*}

\textbf{Figure \ref{channels_and_optimal_codes}a} shows a noisy channel which maps each input $\xrv=x$ to two possible outputs with equal probability. Our goal is to transmit messages from a non-redundant source \figrefsub{channels_and_optimal_codes}{b}, in which three colors occur with equal probability through a noisy channel, with minimal information loss.

\textbf{Figure \ref{channels_and_optimal_codes}c} shows a bad way of doing this. The three chosen states in $\mathcal{X}$ have overlapping outputs in $\yrv$, so information is lost and the original source message cannot always be recovered.

Alternatively, we could design an encoder which uses only a subset of the states in $\mathcal{X}$ that produce disjoint outputs in $\mathcal{Y}$ \figrefsub{channels_and_optimal_codes}{d}. In this case, there is no loss of information.

This encoder is not unique, since we can randomly permute which state of $\mathrm{S}$ mapped to which state of $\mathrm{X}$ and achieve the same mutual information.

\subsubsection{Compatibility of channels and sources}
\label{source_channel_match_sec}

Suppose we are given some arbitrary source, we have multiple channels to choose from, and we can design any type of encoder we want, provided that it only encodes 1 source message at a time. What channel will allow us to transmit the most information? How much information will we be able to transmit? Will this channel also be the best for any other different source? 

The answer to the last question is no. Even with freedom to design the encoder as we see fit, some sources and channels are inherently better ``matched'' to one another. Furthermore, for a given source and channel, it may not be possible to achieve the channel capacity when encoding one message at a time.

Intuitively, how well a source and channel are ``matched'' is determined by how well our best encoder maps more probable source outcomes to 1) inputs with less noisy outputs and 2) inputs whose outputs don't overlap with those of other inputs. The process of creating an encoder to perform this matching is also known as \textbf{joint source-channel coding}.

\begin{figure*}[htbp]
\centering
\includegraphics[width=1\linewidth]{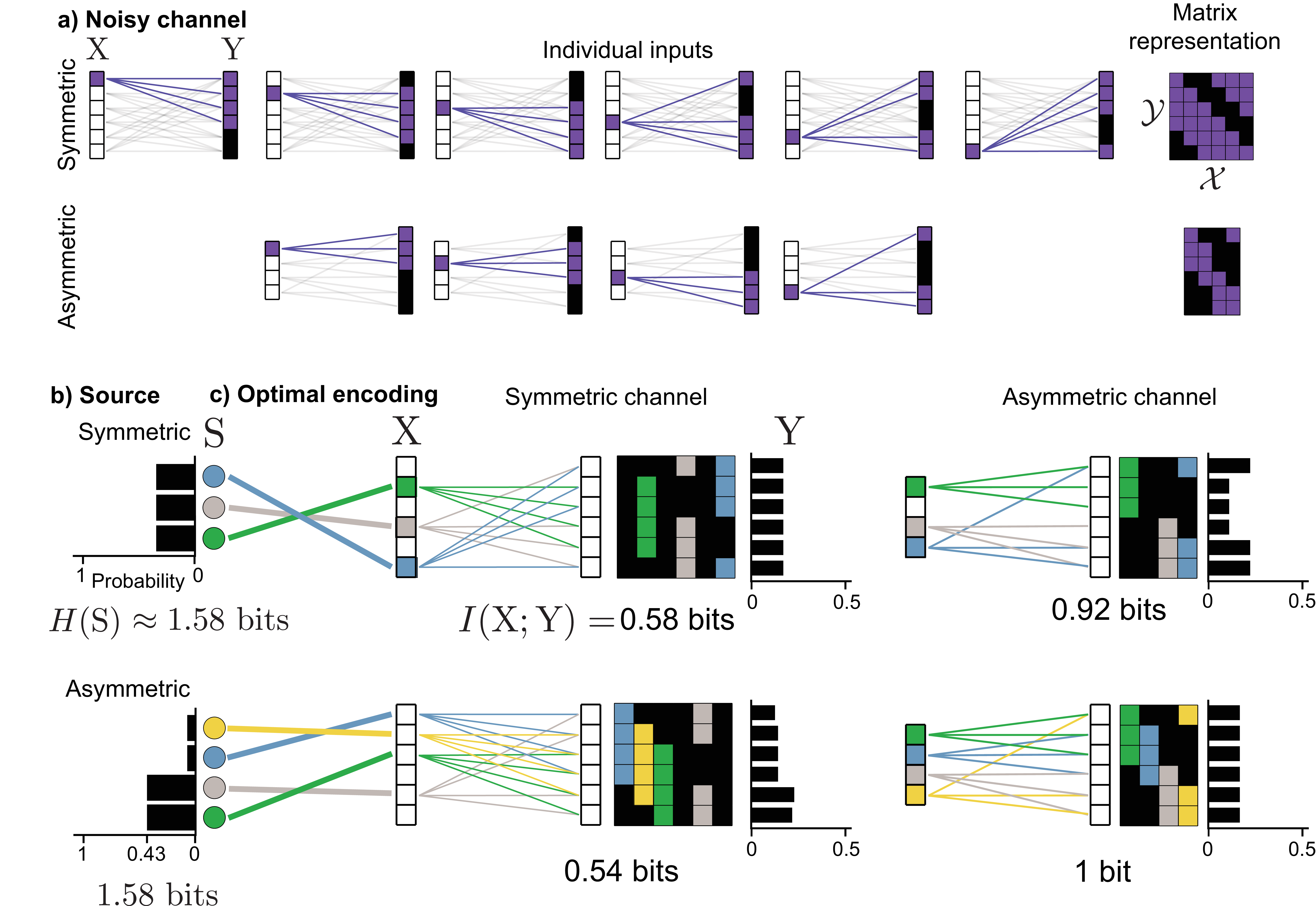}
\caption{\textbf{Matching sources and channels for maximum information transmission.} \textbf{a)} A symmetric noisy channel, in which all inputs are equally noisy and equally overlapping (for a uniform input distribution) and an asymmetric noisy channel, in which all inputs are equally noisy, but two pairs of inputs overlap more at the output with each other than the other pair (for a uniform input distribution). \textbf{b)} Symmetric and asymmetric sources with the same entropy. \textbf{c)}  The symmetric source can be encoded to transmit more information in the symmetric channel, because it is able to choose inputs such that overlap equally at the output, whereas the asymmetric source is not. For the asymmetric channel, this is reversed, and the more probable messages from the asymmetric source can be encoded such that they are less overlapping at the output. }
\label{channel_source_matching}
\end{figure*}

\textbf{Figure \ref{channel_source_matching}} shows an example of the differences in information transmission that occur when matching different source distributions to different channels. There are two noisy channels \figrefsub{channel_source_matching}{a}: The first channel (the ``symmetric channel") has inputs that all have equally noisy outputs that overlap equally. The second channel (the ``asymmetric channel") also has inputs with equally noisy outputs, but those outputs are not equally overlapping with one another: it has two pairs of inputs whose outputs overlap less with each other than do the outputs within each pair.

There are also two different sources with the same entropy \figrefsub{channel_source_matching}{b}. The first source (the ``symmetric source") has three equally probable outcomes, while the second source (the ``asymmetric source") has two more probable and two less probable outcomes. For channels and sources with a small number of states such as the ones shown here, the optimal encoder can be found by exhaustively searching all possible ways of mapping messages to channel inputs. 

For the symmetric channel, the optimal encoder for the symmetric source is able to transmit more information than the optimal encoder for the asymmetric source, but for the asymmetric channel, the reverse is true \figrefsub{channel_source_matching}{c}. In the latter case, this occurs because the encoder is able to map each of source's two most highly probable messages ($\texttt{\color{OliveGreen}{green}}$ and $\texttt{\color{gray}{gray}}$) to distinct subsets of channel inputs that overlap less at the output. This is not possible for the symmetric source, because all of its messages are equally probable.

In contrast, for the symmetric channel, the symmetric source transmits more information than the asymmetric source. Because all inputs have equally overlapping outputs, the asymmetric source cannot exploit its unequal message probabilities — there are no less-overlapping inputs to preferentially assign its most probable messages to.

\paragraph{Maximal information transmission}

We have seen that the channel capacity $C$ is determined by the optimal input distribution, and that encoders mapping one message at a time may not reach this capacity. A central result in information theory, the \textbf{Noisy channel coding theorem}, proves that it is possible to transmit information up to the channel capacity for any source whose entropy rate does not exceed the channel capacity. The key is encoding not one message at a time, but many messages at once. This changes both the distribution of source messages and the noise per channel input, making them more uniform and easier to match with an encoder.

\subsection{The noisy channel coding theorem}

\label{noisy_channel_coding_section}
Noisy channel coding is one of the central problems in information theory. The \textbf{noisy channel coding theorem} defines the limits of reliable transmission and storage of information using inherently noisy physical components.

The general setup is: There is a source of random messages, which could be human-chosen or the result of some natural random process.  The goal is to transmit these messages to a receiver with minimal or no error, but to do so they must pass through a \textbf{noisy channel}.

The noisy channel coding theorem shows that reliable transmission is achievable at any \textbf{rate} (i.e. information per channel use) below the channel capacity, enabling a decoder to infer the true message with arbitrarily small probability of error. It also shows that at rates above the channel capacity, achieving arbitrarily small probability of error is impossible. Combined with source coding (Sec. \ref{source_coding_limit}), this means that any source whose entropy rate does not exceed the channel capacity can be transmitted reliably.

\begin{figure*}[htbp]
\centering
\includegraphics[width=0.75\linewidth]{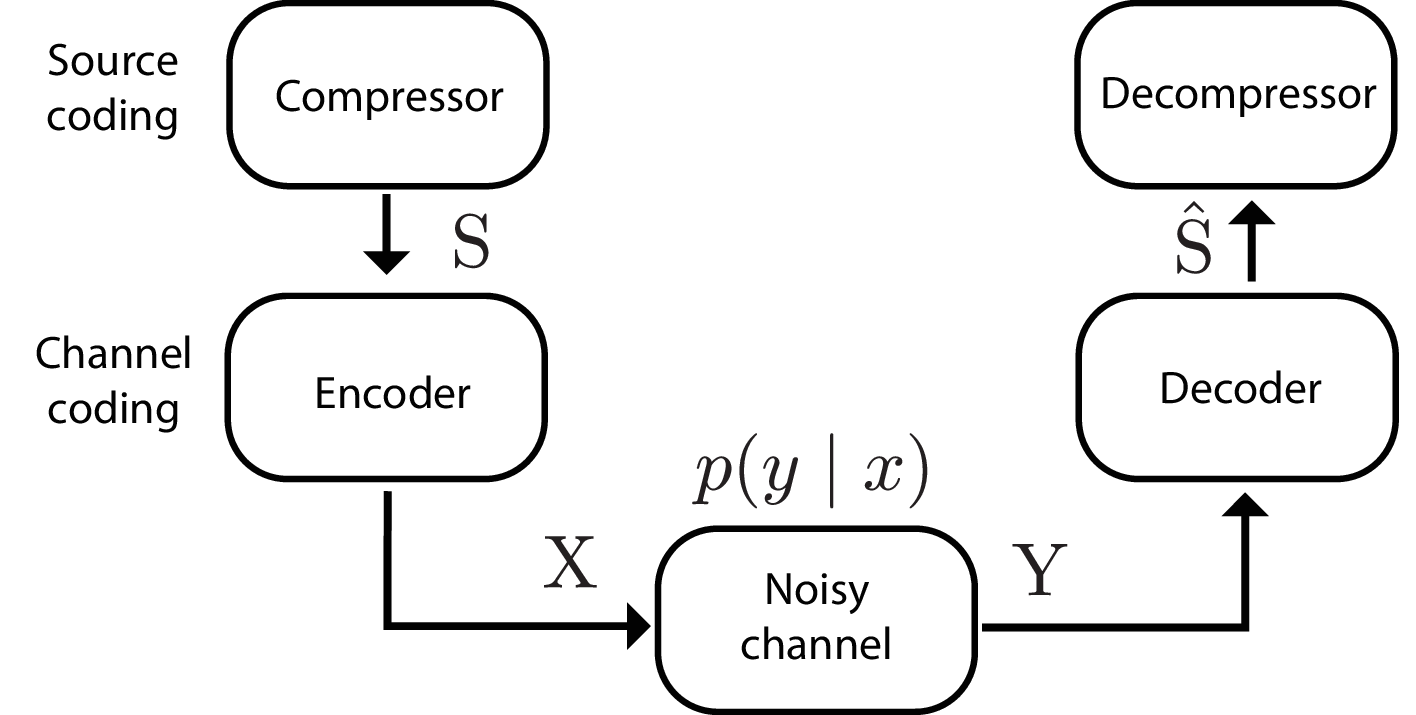}
\caption{\textbf{Noisy channel coding: the big picture} A redundant source passes into a compressor, which yields a compressed  (i.e. maximum entropy) random message $\mathrm{S}$. $\mathrm{S}$ then passes into an encoder, which adds redundancy to create robustness to passage over a noisy channel. The encoded message $\xrv$ passes through the channel, yielding the received message $\yrv$, which possibly contains errors. $\yrv$ then goes into a decoder, yielding an estimate of $\mathrm{S}$, $\mathrm{\hat{S}}$, and finally a decompressor to give an estimate of the source. Adapted from \cite{MacKay2005} p146.}
\label{source_channel_coding}
\end{figure*}

The analysis and proof of the noisy channel coding theorem break the problem into separate source coding and channel coding steps. This yields five steps: \figref{source_channel_coding}: 1) A compressor takes in the random events, and removes any redundancy, making it a maximum entropy source. 2) The compressed events pass into an encoder, which adds redundancy to try to prevent the loss of information when 3) passing through a noisy channel. 4) The channel output is then fed into a decoder, which attempts to remove any errors introduced by the noisy channel and recover the original message. 5) The compressed messages are decompressed back onto the original outcome space.

Steps 2-4 correspond to the random variables:

\begin{itemize}
    \item $S$: The (compressed) source message
    \item $\xrv$: The input to the channel (the encoded message)
    \item $\yrv$: The output of the channel (the received message)
    \item $\hat{S}$: The estimate of the (compressed) message
\end{itemize}

\subsubsection{Ignoring compression with block encoders}
\label{source_channel_sep_section}
The noisy channel coding theorem achieves its result by encoding multiple messages at a time rather than one at a time. The number of messages encoded together is called the \textbf{block length}. In practice, an infinitely long block length cannot be used, because the message would never actually transmit. As a result, a noisy channel might not be able to transmit information up to its full capacity for certain sources. There may also be possible performance gains in practice by using joint source-channel coding, as will be discussed in Section \ref{Noisy_channels_in_practice}.

We separate compression and encoding (and their inverses) into distinct steps not out of necessity, but for easier analysis. The \textbf{source-channel separation theorem} states that there is no cost to this division: compressors and encoders can be designed just as well separately as they can jointly (in the limit of an infinitely long block length).

\begin{figure*}[htbp]
\centering
\includegraphics[width=0.8\linewidth]{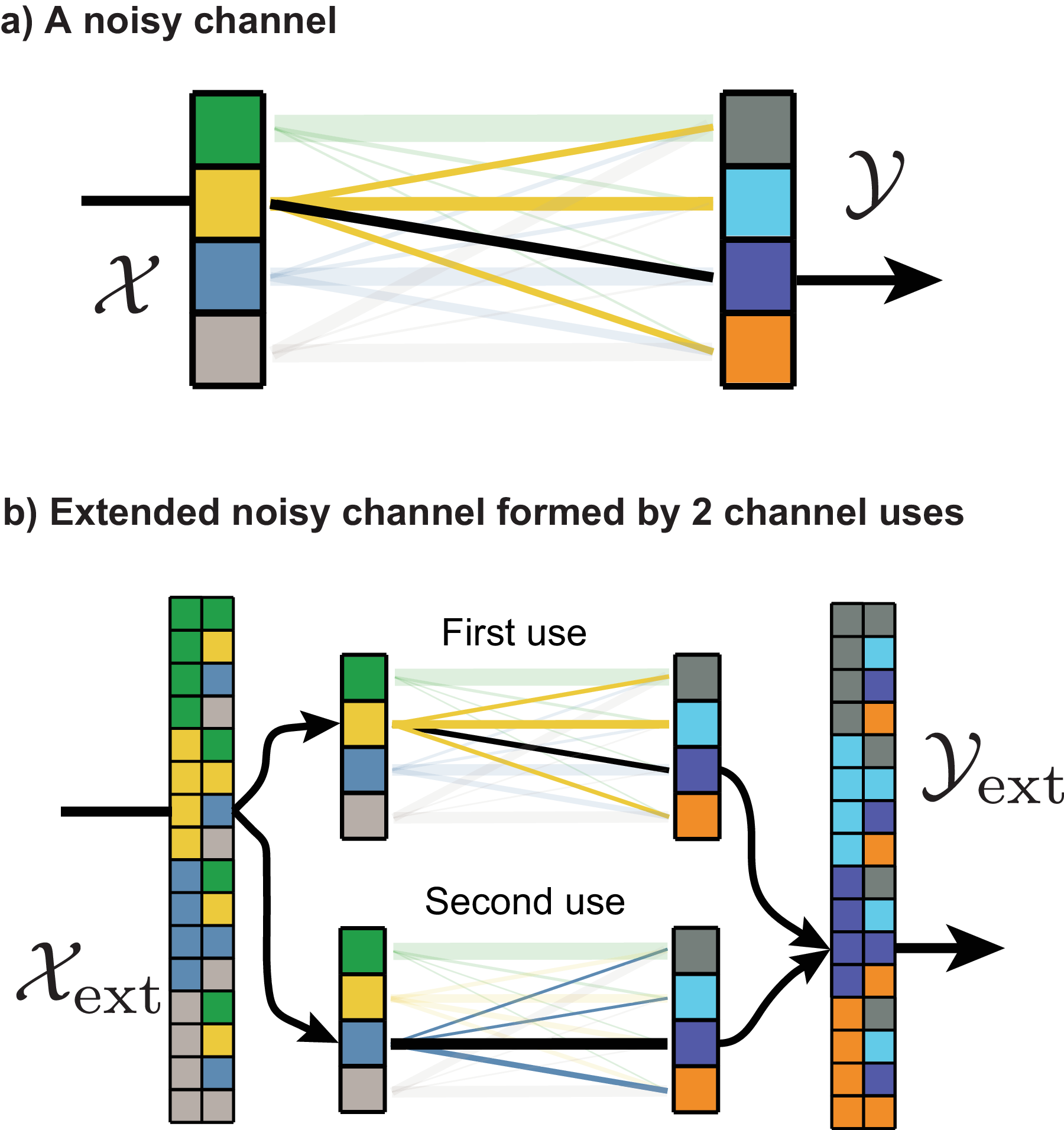}
\caption{\textbf{An extended noisy channel} is formed by treating multiple uses of a single channel as a channel itself. \textbf{a)} One use of the original channel. \textbf{b)} one use of the extended channel. }
\label{extended_noisy_channel_fig}
\end{figure*}

Without loss of generality, we can treat $\mathrm{S}$ as a maximum entropy source and ignore the compression/decompression step. By block-encoding multiple symbols together, any redundant source can be losslessly compressed into a maximum entropy source. This requires a channel with a correspondingly larger state space, which we can form by using an \textbf{extended noisy channel}. This is simply using the channel $N$ times, and considering a meta-channel whose inputs and outputs are all possible combinations of the original channel's inputs and outputs \figref{extended_noisy_channel_fig}.

As a result of this separation and the theorem justifying it, we can disregard the source coding problem (i.e. the compressor and decompressor) and focus only on the channel coding problem (\textbf{encoder $\rightarrow$ noisy channel $\rightarrow$ decoder}). Since any source of random events can be compressed into a maximum entropy source (Sec. \ref{source_coding_limit}), we can simply assume this has already taken place.

\subsubsection{Theoretical limits of perfect transmission}
\label{transmission_limits}

The \textbf{noisy channel coding theorem} proved something previously thought impossible: that the probability of error when transmitting a message over a noisy channel and trying to infer its value on the other side could be made arbitrarily small without the rate of information transfer dropping to zero.

\begin{figure*}[htbp]
\centering
\includegraphics[width=0.95\linewidth]{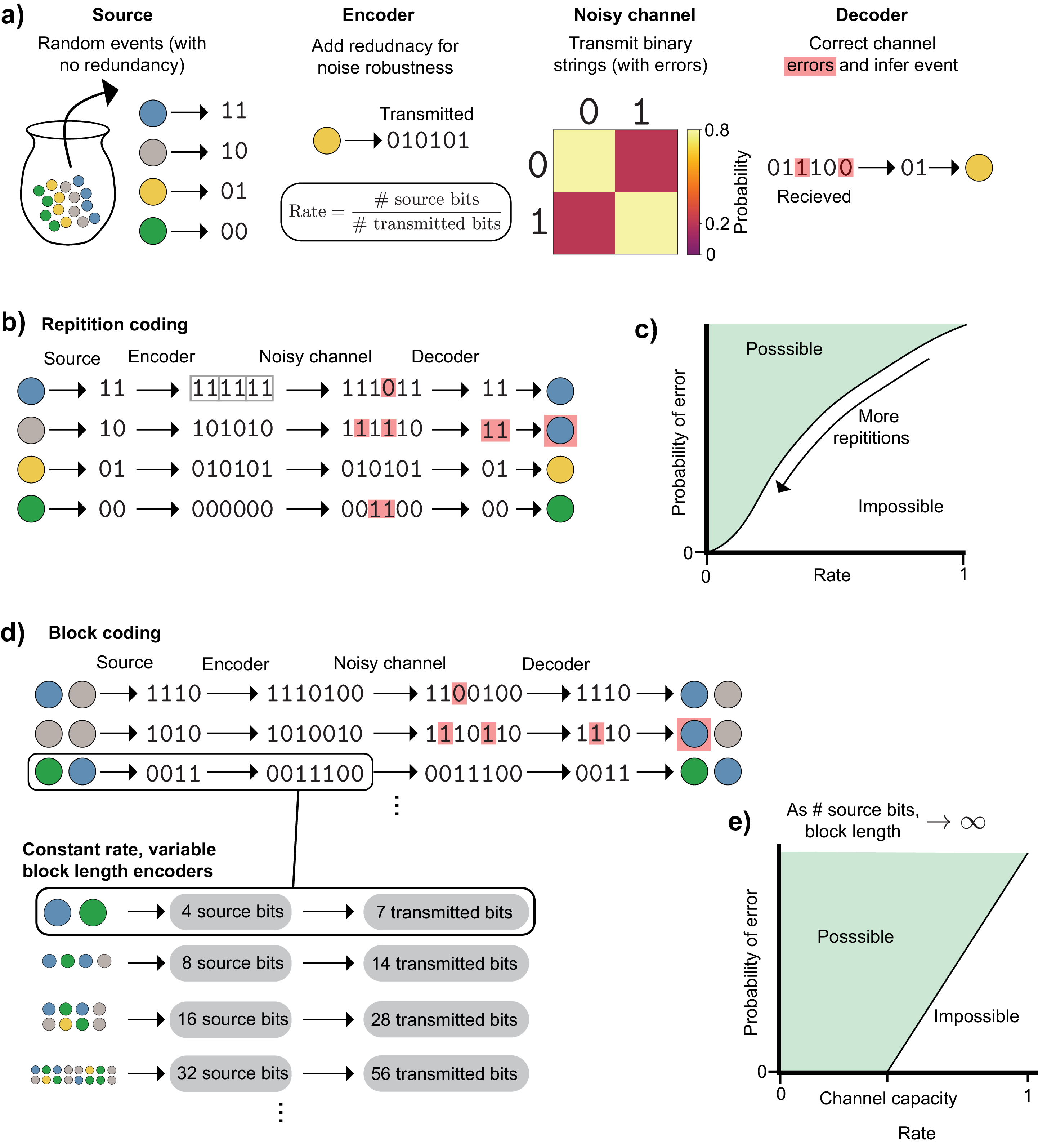}
\caption{\textbf{The noisy channel coding theorem} \textbf{a)} An overview of the problem setup. A source produces non-redundant (i.e. maximum entropy) messages, here a colored marble which is transformed into a two-digit binary encoding. An encoder adds redundancy to prepare a transmitted message (e.g. shown here, repeating the message 3 times). The rate is defined as the ratio of source bits to transmitted bits. The message passes through a noisy channel (e.g. shown here one that flips each bit with probability $\frac{1}{5}$). A decoder attempts to correct errors and recover the original message. \textbf{b)} A repetition coding scheme (same as part a) repeats the source bits a fixed number of times and \textbf{c)} can achieve arbitrarily small probability of error (for the noisy channel in (a)), but needs to send information at a rate approaching $0$ to do so. \textbf{d)} A block coding scheme can do better, by encoding multiple events together. \textbf{e)} As the block length goes to $\infty$, such a scheme can achieve communication with arbitrarily small probability of error at rates up to the channel capacity. At rates above the channel capacity, infinite block lengths will have nonzero probability of error.}
\label{noisy_channel_coding_fig}
\end{figure*}

To understand this, we start with an \textbf{encoder $\rightarrow$ noisy channel $\rightarrow$ decoder}. As shown in \textbf{Figure \ref{noisy_channel_coding_fig}a}, the messages we are trying to send are the outcomes of a series of maximum entropy/non-redundant random events (like colored marbles drawn from an urn with equal probability). The noisy channel transmits binary data, so we will then convert the outcome of each random event to a binary string using an optimal encoder that minimizes the number of bits used. Using binary channels simplifies the analysis, but is not strictly needed. Noisy channels can be over any outcome space, continuous or discrete.

These binary strings will be transmitted over a noisy channel, which will, with some probability, flip $\texttt{0}$s to $\texttt{1}$s and vice versa. If we sent the raw binary messages over this channel, errors would occur and the message on the other side would be interpreted incorrectly, because noise displaces and destroys some of the usable information in each string. To combat this, before transmitting over the noisy channel, we will add some type of redundancy so that the messages are more robust to the channel's noise. Adding this redundancy means that the binary strings transmitted will become longer, but the information they contain will be unchanged. Thus, the number of (information) bits will be smaller than the number transmitted (data) bits.

The \textbf{rate}\footnote{This rate is not exactly the same as the rate used in rate-distortion theory (Sec. \ref{rate_distortion_sec}). Rate as used here is the ratio of source bits to transmitted bits, whereas in rate-distortion theory it describes the absolute amount of information about a particular source.} $R$ at which we transmit data depends on the amount of redundancy added:

\begin{align*}
    R = \frac{\#\text{ source bits}}{\#\text{ transmitted bits}}
\end{align*}

Next, the redundant message passes through a noisy channel. In the example of \textbf{Figure \ref{noisy_channel_coding_fig}a}, this is a \textbf{binary symmetric channel} in which $\texttt{0}$s and $\texttt{1}$s transmit correctly with probability $\frac{4}{5}$ and flip with probability $\frac{1}{5}$.

Finally, a decoder interprets the received message and attempts to correct any errors and reconstruct the original message.

A naive approach to this problem is through repetition coding \figrefsub{noisy_channel_coding_fig}{b}. Here, the message is repeated $N$ times before being sent through the channel, and the decoder chooses the bit at each position that occurred the greatest number of times in the repetition-coded message. The wrong message can be received if more than half of the bits in a given position flip. To lower the probability of this happening, we can increase the number of repetitions. This reduces the probability of error, but in doing so lengthens the message needed for each event and thus lowers the rate at which information can be transmitted. As we specify a lower and lower maximal acceptable probability of message error, the rate at which information is sent approaches $0$ \figrefsub{noisy_channel_coding_fig}{c}.

The noisy channel coding theorem proved that we can do better than this. To do so, we must send multiple messages at once, also known as \textbf{block coding}. For this binary channel, this is done by taking multiple binary messages and passing them through an encoder that maps to a single binary string. The encoder in \textbf{Figure \ref{noisy_channel_coding_fig}d} is called a Hamming code, which is a good choice for this problem, but not specifically needed for the noisy channel coding theorem. As before, the bits are then transmitted and the decoder attempts to correct errors and reconstruct the original messages.

As the bottom panel of \textbf{Figure \ref{noisy_channel_coding_fig}d} shows, by transmitting a sequence of random events as a single outcome, we can increase the block length of our transmission system while leaving the rate unchanged by keeping the ratio of source bits and transmitted bits constant. 

The noisy channel coding theorem considers the performance of channels as the block length goes to infinity, and it has two important implications \figrefsub{noisy_channel_coding_fig}{e}: 
\begin{enumerate}
    \item Encoder/decoder pairs exist which can transmit messages with arbitrarily small probability of error at nonzero rates up to the \textbf{channel capacity}
    \item At rates above the channel capacity, no encoder/decoder pairs can transmit messages with arbitrarily small error probability
\end{enumerate}

The theorem does not state how one can find the appropriate encoders/decoders to achieve this performance, only that they exist. Specifically, it shows that the \textit{average} performance of randomly constructed encoders achieves this, which implies that at least one of the individual randomly constructed encoders can achieve it. The theorem alone does not reveal how to construct good codes with efficient decoders, in part because extending the block length makes naive decoding exponentially more complex. It took several decades of research to discover codes for binary channels that approach the performance the theorem implies.

\subsubsection{The benefits of long block length}

The noisy channel coding theorem shows that information can be transmitted at a rate up to the channel capacity in the asymptotic setting of infinitely long block lengths. This leads to the question: what is it about long block lengths that makes this achievable, even for channel/source pairs (like the ones in \textbf{Figure \ref{channel_source_matching}}) that cannot reach the channel capacity in the short block length setting?

The answer is that long block lengths make both channels and sources more uniform in such a way that the problem of finding an encoder that optimally maps source messages to channel inputs becomes trivial: it can be accomplished by simply picking a fixed random mapping. As discussed in Section \ref{source_channel_match_sec}, a good encoder will map the most probable source messages to channel inputs that are the least noisy (minimize $H(\yrv \mid \xrv)$) as well as choosing inputs that minimize the overlap of messages at the channel output (minimize $H(\xrv \mid \yrv)$). 

By extending a source to a long block length (i.e. encoding a sequence of messages rather than individual messages), all sources will start to resemble maximum entropy sources, which have uniform probability over all messages. This happens either because: 1) individual messages are already coming from a maximum entropy source, and this will remain true at any block length. 2) Redundant sources will produce equally probable typical sequences (discussed in Section \ref{typicality_sec}) onto which nearly all of the probability mass will concentrate, and non-typical sequences can be ignored because they contain vanishingly small probability. Thus, a redundant source, when extended, behaves like a maximum entropy source over a smaller number of possible messages (the typical set).

\begin{figure*}[htbp]
\centering
\includegraphics[width=0.8\linewidth]{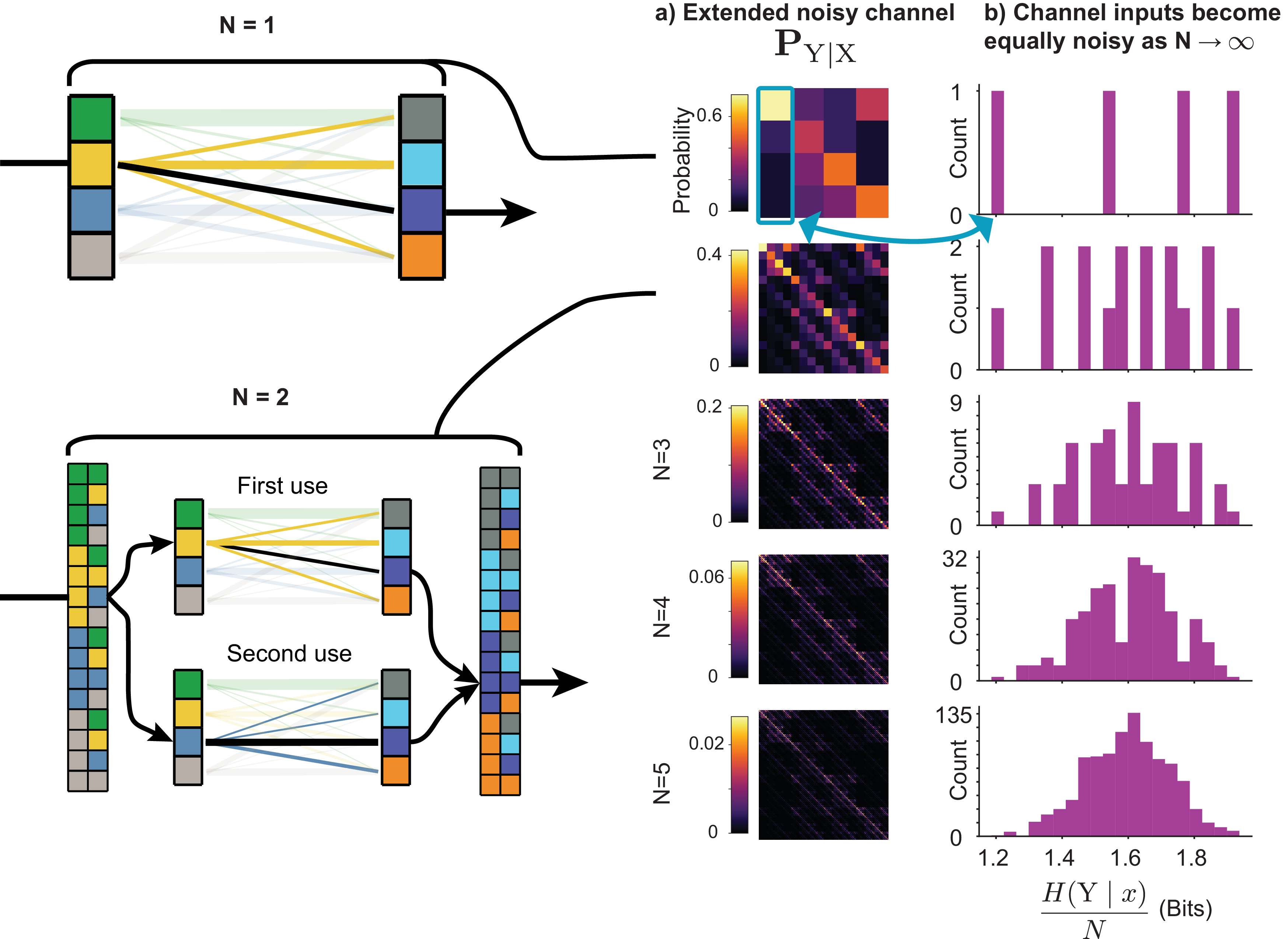}
\caption{\textbf{Conditional entropy in an extended noisy channel.} \textbf{a)} Matrix representation of extended noisy channels with increasing block lengths. \textbf{b)} Histogram of conditional entropy for each channel input. With increasing block lengths, channel inputs tend towards having the same conditional entropy, which means they are equally noisy. }
\label{extended_noisy_channel_cond_entropy_fig}
\end{figure*}

A similar uniformity arises when a noisy channel is extended: both the noise level at each input's output and the overlap of different inputs' outputs (assuming a uniform input distribution) become the same for all inputs. The noisiness of a channel input $x$ is measured by its point-wise conditional entropy $H(\yrv \mid x)$. An input with noiseless outputs would have a value of $0$, and inputs with noisy outputs have positive values. By looking at the distribution of point-wise conditional entropies for each channel input, we can assess how heterogeneous the channel inputs are in terms of their noisiness \figref{extended_noisy_channel_cond_entropy_fig}. The absolute noise level will always increase as we make extended noisy channels with longer block lengths, but the relative amount of noise at each input's output becomes increasingly close to the block length $N$ times the average noisiness of the non-extended channel, $H(\yrv \mid \xrv)$. This is a consequence of the Law of Large Numbers,
as shown in Appendix \ref{ext_channel_proof}.

In addition to this uniformity in the noise level at different channel inputs' outputs, the overlap of different channel inputs' outputs also diminishes as block length increases. To see why, consider that for a given input $x$, the channel's output sequences concentrate on a typical set of size approximately $2^{NH(\yrv \mid x)}$, while the output typical set contains approximately $2^{NH(\yrv)}$ sequences. The fraction of the output space occupied by any single input's outputs is therefore approximately $2^{-N[H(\yrv) - H(\yrv \mid x)]}$. Since extended channel inputs have equally noisy outputs (as shown above), $H(\yrv \mid x) \approx H(\yrv \mid \xrv)$ for most inputs, so this fraction is approximately $2^{-NI(\xrv;\yrv)}$, which shrinks exponentially with $N$. As a result, two randomly chosen inputs become increasingly unlikely to produce overlapping output sets as $N$ grows. This is the intuition behind Shannon's random coding argument. Note that unlike the noise uniformity discussed in the previous paragraph, the overlap depends on a particular choice of input distribution, since the total output space is determined by $\prob{\yrv} = \sum_x \prob{\yrv \mid x}\prob{x}$. However, as $N \rightarrow \infty$, all sources concentrate their probability on typical sequences that are approximately equally probable (Sec. \ref{typicality_sec}), so the input distribution is approximately uniform over the typical set.

Extended noisy channels still have some inputs with especially noisy outputs and some with especially quiet outputs, as well as those that are more or less overlapping. For example, the extended noisy channel input that corresponds to using the input with the noisiest outputs from the non-extended channel for each transmission still may have much noisier outputs than the average extended channel input. Outliers like this don't go away as $N \rightarrow \infty$; they simply cease to matter because they are outnumbered.

However, since we can never actually have an infinite block length in practice, in many situations information transmission can do better than a fixed, random encoder, by shifting more probable messages onto inputs with less noisy, less-overlapping outputs. This is possible because when $N$ is not infinite, any redundant source will not produce equally probable messages. How to do this in practice will be explored in the next section.

\subsection{Noisy channel coding in practice}
\label{Noisy_channels_in_practice}

In practice, block length is finite, and the challenge is to design encoders that maximize information transmission within this constraint. Encoding one message at a time may not reach channel capacity (Sec. \ref{source_channel_match_sec}), and extending the block length to infinity is not feasible, even though it guarantees capacity-achieving encoders exist (Sec. \ref{noisy_channel_coding_section}).

\subsubsection{Longer, but not infinite, block length}
The first important question is: how do the asymptotic changes that improve our ability to match channels and sources manifest as $N$ increases? These changes, the transformation of a source into an extended source with uniform probability over its messages and a channel into an extended noisy channel with inputs whose outputs are equally noisy and overlap equally (for a uniform input distribution), are not only possible as $N \rightarrow \infty$. Some source/channel pairs, like the one shown in \textbf{Figure \ref{channels_and_optimal_codes}}, already have this property with a block length of $1$. However, this represents a special case that will usually not be true.

Generally, a redundant source, when extended, will produce messages that are more uniform in probability as $N$ increases (as shown in \textbf{Figure \ref{typicality_fig}}). Similarly, as $N$ increases, different inputs of a noisy channel will produce outputs that are increasingly similar in noise level and overlap (for equally probable inputs).

Performance gains from increasing block length beyond $N=1$ are initially large but diminish rapidly. Block codes exist whose decoding error probability decreases exponentially with block length (\cite{MacKay2005} p171). For a fixed probability of error, the gap between channel capacity and the actual information transmitted decreases proportional to $\frac{1}{\sqrt{N}}$ \cite{Polyanskiy2010}.

\subsubsection{Optimizing encoders}

With finite block length, sources will not produce equally probable messages and channel inputs will not be equivalent. We can exploit this structure by designing encoders that map more probable messages to less noisy, less-overlapping inputs \cite{Polyanskiy2010}. This approach is called \textbf{joint source-channel coding}, in contrast to designing an encoder that assumes a maximum entropy source.

We can do this by solving the following optimization problem:

\begin{align}
\underset{f}{\arg \max} \: I(f(\mathrm{S}); \yrv) \\
\end{align}

where $f(\cdot)$ is the encoder function mapping from the space of possible messages $\mathcal{S}$ to the space of channel inputs $\mathcal{X}$.

When $\mathcal{S}$ and $\mathcal{X}$ are discrete spaces, this can be a challenging optimization problem, because it is not amenable to gradient-based optimization. If we consider this problem in matrix form, $\mathbf{p}_{\mathrm{S}}$ is a vector of probabilities of different messages, and $\mathbf{P}_{\mathrm{f}}$ is a noiseless channel that maps $\mathbf{p}_{\mathrm{S}}$ to $\mathbf{p}_{\xrv}$. Since the channel is noiseless, each column has exactly one entry with probability $1$, and all other entries are $0$, like the channels shown in \textbf{Figure \ref{noisy_noiseless_mappings}a,c}. 

A naive approach to solving this optimization problem is to use a brute force search, in which the mutual information is calculated for every possible encoder matrix. However, this quickly becomes computationally intractable, since the number of possible encoder matrices grows with the factorial of the number of channel inputs.

\textbf{Stochastic encoders} replace the deterministic function $f(\cdot)$ with a conditional probability distribution $p(x \mid s)$. In matrix form, columns can have any number of nonzero entries, so long as they are all positive and each column sums to $1$, like the matrices in \textbf{Figure \ref{noisy_noiseless_mappings}b,d}.

Since every deterministic encoder represents a specific case of a larger class of stochastic encoders, stochastic encoders can do as well or better in principle, though designing good ones can present practical challenges \cite{Theis2021, Wagner2021}.

The major advantage that stochastic encoders do have is that they are much more amenable to optimization. Since each column of a stochastic encoder matrix (or any channel) is a conditional probability distribution it can be optimized with respect to individual probabilities using the same numerical optimization tools used to find the optimal input distribution for a fixed channel (Appendix \ref{optimizing_mutual_info}).

\paragraph{Continuous encoders}

Encoders can also be optimized for continuous sources and channels. That is, a source is represented by a probability density function, and the encoder is a continuous function (if it is deterministic) or a conditional probability density function (if it is stochastic). In this setting, the entropies become differential entropies, which have a less clear intuitive interpretation and can in some cases take on infinite values (Section \ref{differential_entropy_sec}). However, the gradients of these entropies are well defined and do not suffer from such mathematical difficulties, making gradient-based optimization possible \cite{Bell1995}. Still, there remain many difficulties to optimizing mutual information in both continuous and discrete settings, particularly in high-dimensional settings, and this remains an active area of research \cite{poole2019}.

\subsubsection{Variable channels and the cliff effect}

The analysis so far assumes that the channel's noise characteristics are known and fixed. When the actual channel deviates from the one the encoder was designed for, performance can degrade rapidly, a phenomenon known as the \textbf{cliff effect} \cite{Polyanskiy2010}. The noisy channel coding theorem offers no guidance on designing encoders robust to such variation, and this remains an active area of research.

\subsection{Decoders}

Even with an optimal encoder, the original message $\mathrm{S}$ must still be inferred from the noisy channel output $\yrv$. The \textbf{decoder} performs this inference, producing an estimate $\hat{\mathrm{S}}$ of the original source message. Encoders and decoders must be designed together, accounting for the channel's noise characteristics.

The channel output will contain both source information $I(\xrv; \yrv)$ and noise added by the channel $H(\yrv \mid \xrv)$. An optimal decoder attempts to infer the original message from the noisy output, minimizing the probability of error. In practice, our decoder may not fully exploit the signal available in $\yrv$, resulting in $I(\hat{\mathrm{S}}; \mathrm{S}) < I(\yrv; \mathrm{S})$.

Preserving the information contained in the received message $\yrv$ is necessary, but not sufficient, for estimating the correct message. To understand why it is necessary, consider the rate distortion curve in \textbf{Figure \ref{rate_distortion_fig}c}. The probability that the estimated source message is not equal to the actual source message is a valid distortion function, and the rate-distortion curve tells us that to improve the best possible average performance on this distortion function, we will need more information. To understand why it is not sufficient, consider that if we had a perfect decoder, we could always permute all of its predictions such that it would be wrong every time. This would give the worst possible distortion, but all the information would remain, because we could always perform the inverse permutation.

Block encoders with extended noisy channels increase information throughput up to the channel capacity, but they also make decoder design more complex. Specifically, as we increase the block length, the number of different messages we have to decode increases exponentially: $|\mathcal{Y}|^N$. A naive decoder would simply be a lookup table that maps a received message $\mathrm{Y}$ to an estimate of the source event $\mathrm{\hat{S}}$, and this lookup table becomes exponentially large with increasing block length. Good encoder-decoder pairs thus combine random-like code construction with enough structure to permit decoding algorithms that avoid the exponential complexity of the naive approach. There is not a known formula for designing such encoder-decoder pairs that works across different types of channels.

\section{Code availability}

The code used to produce the figures and high-resolution versions of the figures can be found at \url{https://doi.org/10.5281/zenodo.6647779}

\section{Acknowledgements}

We thank Matt Thomson for alerting H.P. to the existence of information theory, and in particular the wonderful textbook \cite{MacKay2005} by David Mackay that made learning accessible and interesting, Chenling Antelope Xu for being an insightful and thorough member of a study group of that book, Tom Courtade for providing feedback and clarifying concepts, and Leyla Kabuli and Tiffany Chien for their extremely helpful feedback in the preparation of this manuscript.

\appendix

\section{Numerical optimization of channel input distribution}

\label{optimizing_mutual_info}

Computing the optimal input distribution for a noisy channel can be used not only to design encoders that provide maximal information throughput, but also to compute the channel's capacity. In some cases we may be able to use analytical properties of the channel (e.g. Gaussian noise with zero mean known variance) to perform this computation. In other cases, we can find this distribution by maximizing the channel's mutual information with respect to the input probabilities (\cite{MacKay2005} p169). Mathematically:

\begin{align}
\mathbf{p}_{\xrv}^* = \underset{\mathcal{P}_{\mathcal{X}}}{\arg \max} \: I(\xrv; \yrv) \\
C = \underset{\mathcal{P}_{\mathcal{X}}}{\max} \: I(\xrv; \yrv) \\
\end{align}

This optimization problem can be solved, and the optimal distribution/channel capacity computed, using numerical optimization techniques such as gradient ascent. Mutual information is a concave function of the input probabilities for a fixed channel, so gradient ascent over the input probabilities is guaranteed to find the global maximum with a proper learning rate. 

We solve this problem using projected ascent, which consists of applying the following updates:

\begin{align*}
    \mathbf{p}_{\xrv}^{k+1} =  \text{proj}(\mathbf{p}_{\xrv}^{k} + \lambda \nabla_{\mathbf{p}_{\xrv}} I(\xrv; \yrv))
\end{align*}

where $\mathbf{p}_{\xrv}$ is a vector of probabilities for each state in $\mathcal{X}$.

The projection operator $\text{proj}(\cdot)$ approximately enforces the constraints that the probabilities need to be positive and sum to 1. For a sufficiently small step size $\lambda$, this can be done by taking the element-wise maximum of $\mathbf{p}_{\xrv}$ and $\mathbf{0}$ (a vector of all $0$s), and then adding enough to each element such that they sum to one. Mathematically:

\begin{align*}
    \mathbf{p}_{\xrv} &= \max(\mathbf{p}_{\xrv}, \mathbf{0}) \\
    \mathbf{p}_{\xrv} &=  \mathbf{p}_{\xrv} + \frac{1 - \sum_{i=1}^{N} p_i}{N}
\end{align*}

where $N$ is the number of elements in $\mathbf{p}_{\xrv}$ and $p_i$ is the $i$th entry of $\mathbf{p}_{\xrv}$.

While this heuristic projection step appears to work in practice, we note that there may be better and more theoretically motivated alternatives \cite{Wang2013a}.

\section{Proofs}
\subsection{Proof that joint entropy of two independent random variables equals sum of individual entropies}
\label{independent_entropies_proof}
\begin{align*}
    H(\xrv, \yrv) &= \sum_{x \in \mathcal{X}} \sum_{y \in \mathcal{Y}} p(x, y) \log{\frac{1}{p(x, y)}} \\
                &= \sum_{x \in \mathcal{X}} \sum_{y \in \mathcal{Y}} p(x)p(y) \log{\frac{1}{p(x)p(y)}} \\
                &= \sum_{x \in \mathcal{X}} \sum_{y \in \mathcal{Y}} p(x)p(y) \left(\log{\frac{1}{p(x)}} + \log{\frac{1}{p(y)}}\right) \\
                &= \left(\sum_{x \in \mathcal{X}} \sum_{y \in \mathcal{Y}} p(x)p(y) \log{\frac{1}{p(x)}}\right) +
                 \left(\sum_{x \in \mathcal{X}} \sum_{y \in \mathcal{Y}} p(x)p(y) \log{\frac{1}{p(y)}} \right) \\
                &= \left(\sum_{x \in \mathcal{X}}  p(x) \log{\frac{1}{p(x)}} \sum_{y \in \mathcal{Y}}p(y)\right) +
                 \left( \sum_{y \in \mathcal{Y}} p(y) \log{\frac{1}{p(y)}} \sum_{x \in \mathcal{X}}p(x) \right)
\end{align*}

Taking advantage of the fact that the sum over any probability distribution is one, this reduces to:

\begin{align*}
     &= \left(\sum_{x \in \mathcal{X}}  p(x) \log{\frac{1}{p(x)}}\right) +
     \left( \sum_{y \in \mathcal{Y}} p(y) \log{\frac{1}{p(y)}} \right) \\
     &= H(\xrv) + H(\yrv)
\end{align*}

\subsection{Decomposition of mutual information}
\label{mi_decomp_proof}

\begin{align*}
        I(\xrv; \yrv) &= \sum_{x \in \mathcal{X}} \sum_{y \in \mathcal{Y}} p(x, y)\log \left( \frac{p(x, y)}{p(x)p(y)} \right) \\
        &= \sum_{x \in \mathcal{X}} \sum_{y \in \mathcal{Y}} p(x, y)\log \left( \frac{p(y \mid x)}{p(y)} \right) \\
        &= \sum_{x \in \mathcal{X}} \sum_{y \in \mathcal{Y}} p(x, y)(- \log p(y) + \log p(y \mid x)  )  \\
        &= -\left( \sum_{x \in \mathcal{X}} \sum_{y \in \mathcal{Y}} p(x, y)\log  p(y) \right) + 
        \left( \sum_{x \in \mathcal{X}} \sum_{y \in \mathcal{Y}} p(x, y)\log  p(y \mid x) \right)  \\
        &= - \left(  \sum_{y \in \mathcal{Y}} \left(\sum_{x \in \mathcal{X}} p(x, y)\right) \log  p(y)  \right) + 
        \left( \sum_{x \in \mathcal{X}} \sum_{y \in \mathcal{Y}} p(x, y)\log  p(y \mid x) \right) \\
        &= - \left(  \sum_{y \in \mathcal{Y}} p(y) \log  p(y) \right) + 
           \left( \sum_{x \in \mathcal{X}} \sum_{y \in \mathcal{Y}} p(x, y)\log  p(y \mid x) \right) \\
        &= H(\yrv) - H(\yrv \mid \xrv)
\end{align*}

\subsection{Proof that block length-normalized extended noisy channel inputs become equally noisy for infinite block length}
\label{ext_channel_proof}

Since each channel use within the extended noisy channel is independent, the point-wise conditional entropies will add. For a sequence of channel uses where $x^{(k)}$ represent the input used on the $k$th transmission of the non-extended channel:

\begin{align}
    H(\yrv \mid \mathbf{x}) &= \\
    &= H(\yrv^{(1)}, \yrv^{(2)},...,\yrv^{(N)} \mid x^{(1)}, x^{(2)},..., x^{(N)}) \\
    &= H(\yrv^{(1)} \mid x^{(1)}) + H(\yrv^{(2)} \mid x^{(2)}) + ... + H(\yrv^{(N)} \mid x^{(N)})
\end{align}

where $\mathbf{x} = x^{(1)}, x^{(2)}, ...$ represents the input used on each channel use. These too are independent random variables, since the input chosen on one channel use is independent of subsequent channel uses. Thus, we can take a probability-weighted average over the choice of channel input, yielding:

\begin{align*}
    \sum_{\mathbf{x} \in \mathbf{\mathcal{X}^N}} p(\mathbf{x}) H(\yrv \mid \mathbf{x}) 
    &= \left(\sum_{\mathbf{x} \in \mathbf{\mathcal{X}^N}} p(\mathbf{x}) H(\yrv \mid x^{(1)}) \right) + \left(\sum_{\mathbf{x} \in \mathbf{\mathcal{X}^N}} p(\mathbf{x})H(\yrv \mid x^{(2)})\right) + ... \\
    &= \left(\sum_{x^{(1)} \in \mathcal{X}} p(x^{(1)}) H(\yrv \mid x^{(1)}) 
    \left[\prod_{i=2}^{N}\sum_{x^{(i)} \in \mathcal{X}} p(x^{(i)}) \right]\right) +  \\
    & \left(\sum_{x^{(2)} \in \mathcal{X}} p(x^{(2)}) H(\yrv \mid x^{(2)}) 
    \left[\sum_{x^{(1)} \in \mathcal{X}}p(x^{(1)})\prod_{i=3}^{N}\sum_{x^{(i)} \in \mathcal{X}} p(x^{(i)}) \right]\right) ...
\end{align*}

By using the fact the sum of a probability distribution is equal to one, this reduces to:

\begin{align*}
     & \left(\sum_{x^{(1)} \in \mathcal{X}} p(x^{(1)}) H(\yrv \mid x^{(1)})\right) + 
     \left(\sum_{x^{(2)} \in \mathcal{X}} p(x^{(2)}) H(\yrv \mid x^{(2)})\right) + ... \\
     &= H(\yrv \mid \xrv^{(1)}) + H(\yrv \mid \xrv^{(2)}) + ... \\
     &= \sum_{k=1}^N H(\yrv \mid \xrv)
\end{align*}

Since each term in this sum is identical, dividing by $N$ gives exactly $H(\yrv \mid \xrv)$: the expected block length-normalized noise across extended channel inputs equals the average noise of the non-extended channel. Since the block length-normalized noise of an individual input $\frac{1}{N}\sum_{k=1}^{N} H(\yrv \mid x^{(k)})$ is a sample mean of independent terms, the Law of Large Numbers guarantees that it concentrates around $H(\yrv \mid \xrv)$ as $N \rightarrow \infty$.

\bibliographystyle{unsrt}
\bibliography{references}

\end{document}